\documentclass[12pt]{article}
\usepackage[latin1]{inputenc}
\usepackage[dvips]{graphicx}
\usepackage{t1enc}
\usepackage{dcolumn}
\usepackage[table]{xcolor}
\usepackage{color}
\usepackage{float}
\usepackage{tabularx}
\usepackage{booktabs}
\usepackage{caption}
\usepackage{lscape}
\usepackage{setspace}
\usepackage{lscape}
\usepackage{amsmath,longtable,multicol,dcolumn,tabularx,graphicx,amssymb}
\usepackage{exscale,amsthm,amssymb}
\usepackage{epstopdf}
\usepackage{comment}
\usepackage{dsfont}
\begin{document}


\renewcommand{\baselinestretch}{1.0}
\newcolumntype{.}{D{.}{.}{-1}}

\title{Analyzing  Commodity Futures Using Factor State-Space Models with
Wishart Stochastic Volatility}
\author{Tore S.~Kleppe\\ {\em  Department of Mathematics and Physics, University of Stavanger, Norway}\\[0.0cm]  Roman Liesenfeld\thanks{Corresponding address: Institute of Econometrics and Statistics, University of Cologne,
Albertus-Magnus-Platz, D-50937 Cologne, Germany.  Tel.: +49(0)221-470-2813;
fax: +49(0)221-470-5074. {\sl E-mail address:}
liesenfeld@statistik.uni-koeln.de}\\ {\em  Institute  of Econometrics  and Statistics, University of Cologne, Germany}\\[0.0cm]
Guilherme Valle Moura\\ {\em  Department of Economics,  Federal University of Santa Catarina, Brasil}\\[0.0cm]
Atle Oglend\\ {\em  Department of Industrial Economics, University of Stavanger, Norway}\\[0.0cm]}
\date{\today}

\newcommand{\bs}{\boldsymbol}

\date{(August 20, 2019)}

\maketitle 
\renewcommand{\baselinestretch}{1.5}
\large
\normalsize
\vspace{-1cm}
\begin{abstract}
We propose a factor state-space approach with stochastic volatility to model and forecast the term structure of future contracts on commodities. Our approach builds upon the dynamic 3-factor Nelson-Siegel model and its 4-factor Svensson extension and assumes for the latent level, slope and curvature factors a Gaussian vector autoregression  with a multivariate  Wishart stochastic volatility process.
Exploiting the conjugacy of the Wishart and the Gaussian distribution, we develop a computationally fast and easy to implement  MCMC algorithm for the Bayesian posterior analysis.
An empirical application to daily prices for contracts on crude oil with stipulated delivery dates ranging from one to 24 months ahead show that the estimated 4-factor Svensson model with two curvature factors  provides a good parsimonious representation of the serial correlation in the individual prices and their volatility. It also shows that this model has a good out-of-sample forecast performance.
\end{abstract}
\vspace*{0.2cm} {\sl JEL classification: C32, C38, C51, C58, G13, Q02}

\\


\section{Introduction}
There has been considerable interest in modelling and forecasting commodity futures term structures. Appropriate methods to model term structures, and thus the expected future path of commodity prices, allow better decisions with respect to portfolio holdings, risk management, and dynamic hedging strategies.

In the present paper, we propose a dynamic factor state-space model with Wishart multivariate stochastic volatility (SV) for the term structure of prices for future contracts on crude oil, and develop a computationally fast and easy to implement Markov-Chain Monte-Carlo  (MCMC) algorithm for its Bayesian posterior analysis. We apply our model to  multiple time series consisting of daily prices for 24 monthly future contracts on light sweet crude oil and analyze its ability to represent the joint dynamics in the level and volatility of the prices along the term structure and its out-of-sample forecast performance.

The existing literature on modelling and predicting commodity futures term structures can be classified in terms of three distinct approaches: financial, economic and statistical. The financial literature applies adaptations of affine term-structure models from financial asset pricing theory to model the convenience yield, often defined as the (net) flow of services that accrues to a holder of the physical commodity, but not to a holder of a contract for future delivery (Brennan, 1991). This literature can be traced to the seminal contribution of Black (1976). The standard approach is to impose an exogenous affine jump diffusion process for the convenience yield (and possibly for other pricing factors). The economic approach is based on the competitive storage model of commodity prices, which treats the convenience yield as a real option on storage. Oglend and Kleppe (2019) show how this competitive storage framework can be used to directly model commodity futures term structures, and compares the economic and financial models.

In this paper, we follow the third approach, which treats term structure modelling as a fundamentally statistical problem. This strand of the literature exploits that price data for commodity futures markets and fixed-income markets share a very similar structure, and therefore use parametric reduced-form models originally designed to parsimoniously approximate the time variation in fixed-income term-structure curves (West, 2011; Gr{\o}nberg and Lunde, 2016). This reduced-form approach typically builds upon the dynamic version of the Nelson-Siegel model (NS, Nelson and Siegel, 1987)  introduced by Diebold and Li (2006), which uses three unobserved dynamic factors to represent the level, slope and curvature of the term structure. While less theoretically rigorous, such reduced-form models often show better out-of-sample forecasting performance, and are typically easier to implement empirically than corresponding structural economic/financial approaches.

For the empirical implementation of the dynamic NS approach, Diebold and Li (2006) propose a two-step procedure, which extracts in the first step the factors using cross-section least squares. The second step then consists in applying time series models such as vector autoregressions (VAR) to the extracted factors. However, such an approach suffers from the fact that the statistical inference in the second step ignores estimation uncertainty associated with the factor extraction in the first step. Hence, Diebold et al.~(2006) propose to frame the NS factor structure into a state-space model, where the factors are treated as latent state processes, so that the model can be analyzed by standard Bayesian or classical likelihood procedures. Prominent extensions of this  3-factor NS approach for interest rates include additional factors, either observed ones such as the inflation rate (Diebold et al., 2006), or unobserved ones adding flexibility to the potential shape of the term structure (Svensson, 1994). Other important extensions account for the well-documented volatility clustering in  fixed-income rates through dynamic volatility models either for the volatility of the pricing errors (Koopman et al., 2010 ) or of the factors (Hautsch and Ou, 2012).

Applications of the dynamic  3-factor NS approach for modelling commodity futures are found in West (2011), Gr{\o}nberg and Lunde (2016), Barun\'{\i}k and Mal\'{\i}nsk\'{a} (2016), Etienne and Mattos (2016) and Karstanje et al.~(2017). West (2011) and  Etienne and Mattos (2016) use it to analyze agricultural futures while the contributions of Gr{\o}nberg and Lunde (2016) and  Barun\'{\i}k and Mal\'{\i}nsk\'{a} (2016) consider crude oil futures. Karstanje et al.~(2017) extend the NS model to a joint framework to study comovement across the markets for the most traded commodity futures. The study most closely related to ours is that of Gr{\o}nberg and Lunde (2016) who account for
time-varying volatility in crude oil future prices by modelling the three NS factors using GARCH models in a copula framework. However, since efficient one-step maximum likelihood (ML) estimation for that model is not feasible the authors rely on multistep ML approach based on a factor extraction using the Diebold and Li (2006) procedure.

We make several contributions to this literature on reduced-form modelling of commodity future prices. First, we propose a factor state-space aproach based on the Svensson (1987)  extension of the baseline NS model which obtains by adding to the three NS factor an additional fourth factor for the curvature of the term structure. Although the Svensson model has been used very successfully by practitioners and central banks to model and forecast more complex yield-curve shapes, it has (to the best of our knowledge) not yet been applied for commodity futures.
Second, we account for time-variation in the volatility of future prices through a combination of a Gaussian VAR for the latent factors with a Wishart multivariate SV model for their innovations. This volatility model initially proposed by Uhlig (1994, 1987) for marcoeconomic VAR applications is very parsimonious and easy to handle in a likelihood based analysis, but still  very flexible in representing non-trivial dynamics in covariance matrices. In our application to crude oil futures both the fourth factor as well as the Wishart volatility process proves to be important not only for the in-sample performance of our model but also for its  out-sample predictive fit and and its ablility to predict  the value-at-risk (VaR) of portfolios consisting of future contracts. Third, exploiting the conjugacy  of the Gaussian and Wishart distribution in the latent factor specification we are able to develop  a fast and easy to implement MCMC procedure for a Bayesian posterior analysis of our model. The simplicity of this MCMC procedure  also allows us to easily compute  a wide range of statistics for the assessment of the out-of-sample forecast performance.

\section{Data}
Our data set consists  of daily  settlement prices of 24 monthly future contracts  on light sweet crude oil (WTI) traded at the New York Mercantile Exchange  (until 2008) and the Commodity Exchange (after 2008). The sample period ranges from  Jan 2, 1996 until May 31, 2016 with 5118 price observations for each of the 24 contracts.
Prices are quoted in US-dollar per barrel. The contracts used are  the 24 consecutive closest to delivery  contracts  with stipulated delivery ranging from one to 24 months ahead.
At the end of each month, all contracts are rolled-over to their respective  next nearby monthly contracts.
As such, no contracts in the sample are priced within last trading months preceding delivery. This is a common procedure that avoids problems with declining volume and open interest in these months.
It produces 24  price series for `perpetual' monthly contracts with day-to-day decreases in time to maturities and a maturity increase by a month at roll-over days. At the selected roll-over dates, the time to maturity for the contract closest to delivery ranges from 20 to 24 days.

\begin{center}
\begin{figure}[!htb]
\centering
\includegraphics[width=0.8\textwidth,angle=0]{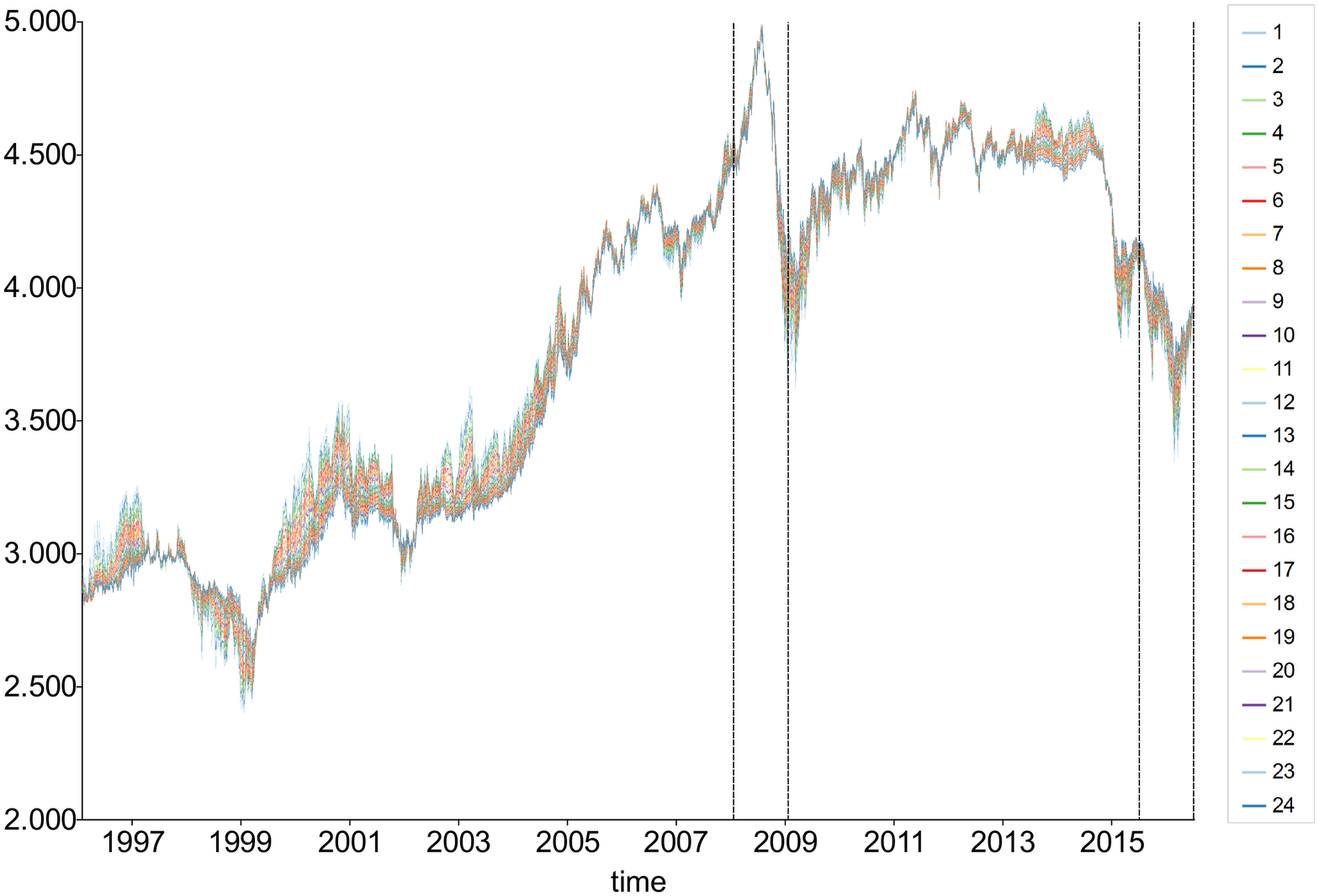}\\[-0.6cm]
\hspace*{0.6cm}\includegraphics[width=0.77\textwidth,angle=0]{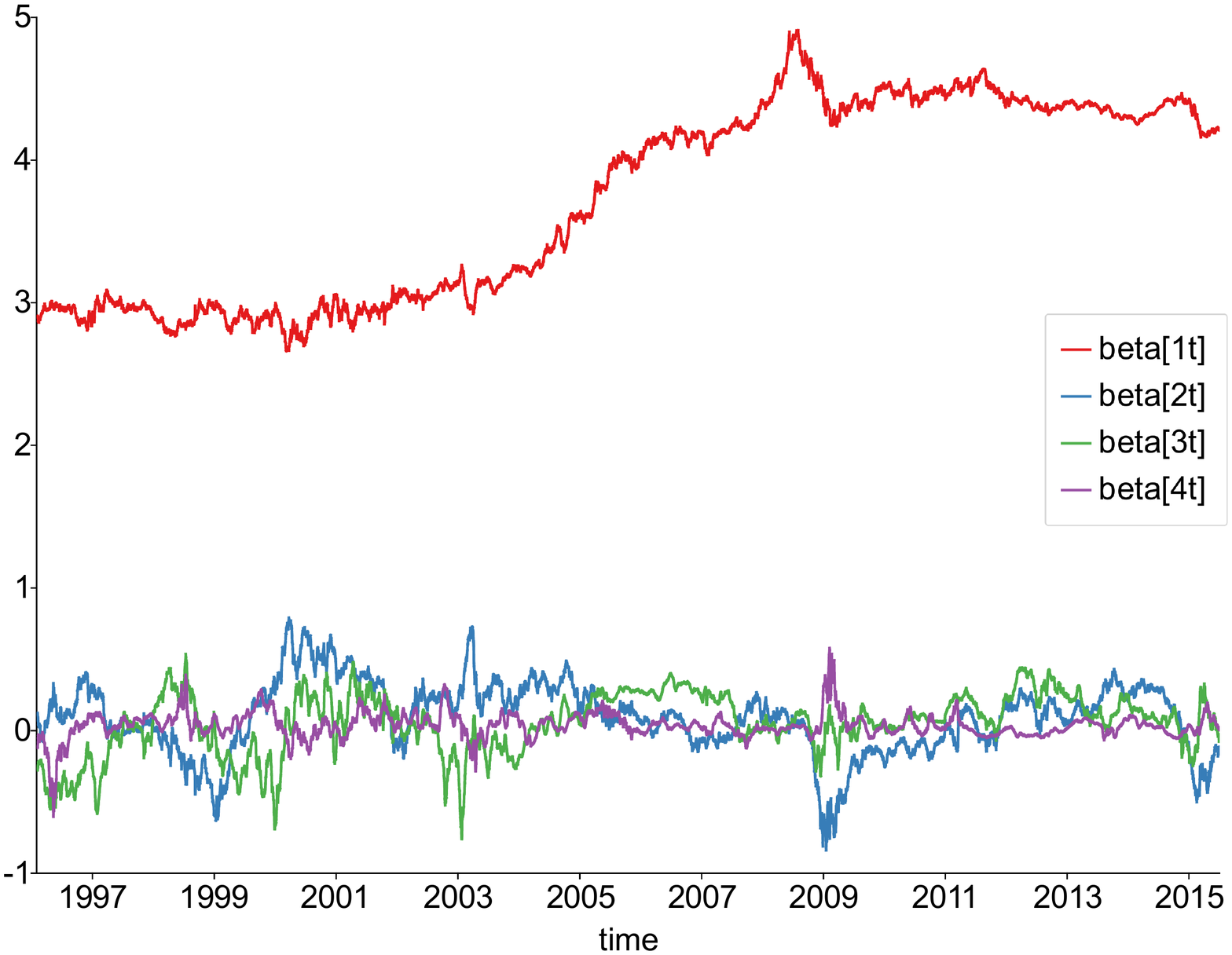}
\captionsetup{font=small}
\caption{Upper panel: Time series plots of the log prices   of the  24 monthly  closest-to-delivery  crude oil future contracts  from  Jan 2, 1996 to May 31, 2016; 'Contract 1 is the next expiring contract and `contract 24'  the last expiring contract; The black vertical lines in the upper panel mark  the two out-of-sample windows  used in the forecasting experiments; The first out-of-sample window ranges from Jan 2, 2008 to Dec 31, 2008 and the second from Jun 1, 2015 to May 31, 2016. Lower panel: Time series plots of the posterior means for the factors $(\beta_{1t},\beta_{2t},\beta_{3t},\beta_{4t})$ obtained under the 4F-SV model.   \label{fig:data}}
\end{figure}
\end{center}

\begin{center}
\begin{figure}[!htb]
\centering
\includegraphics[width=0.8\textwidth,angle=0]{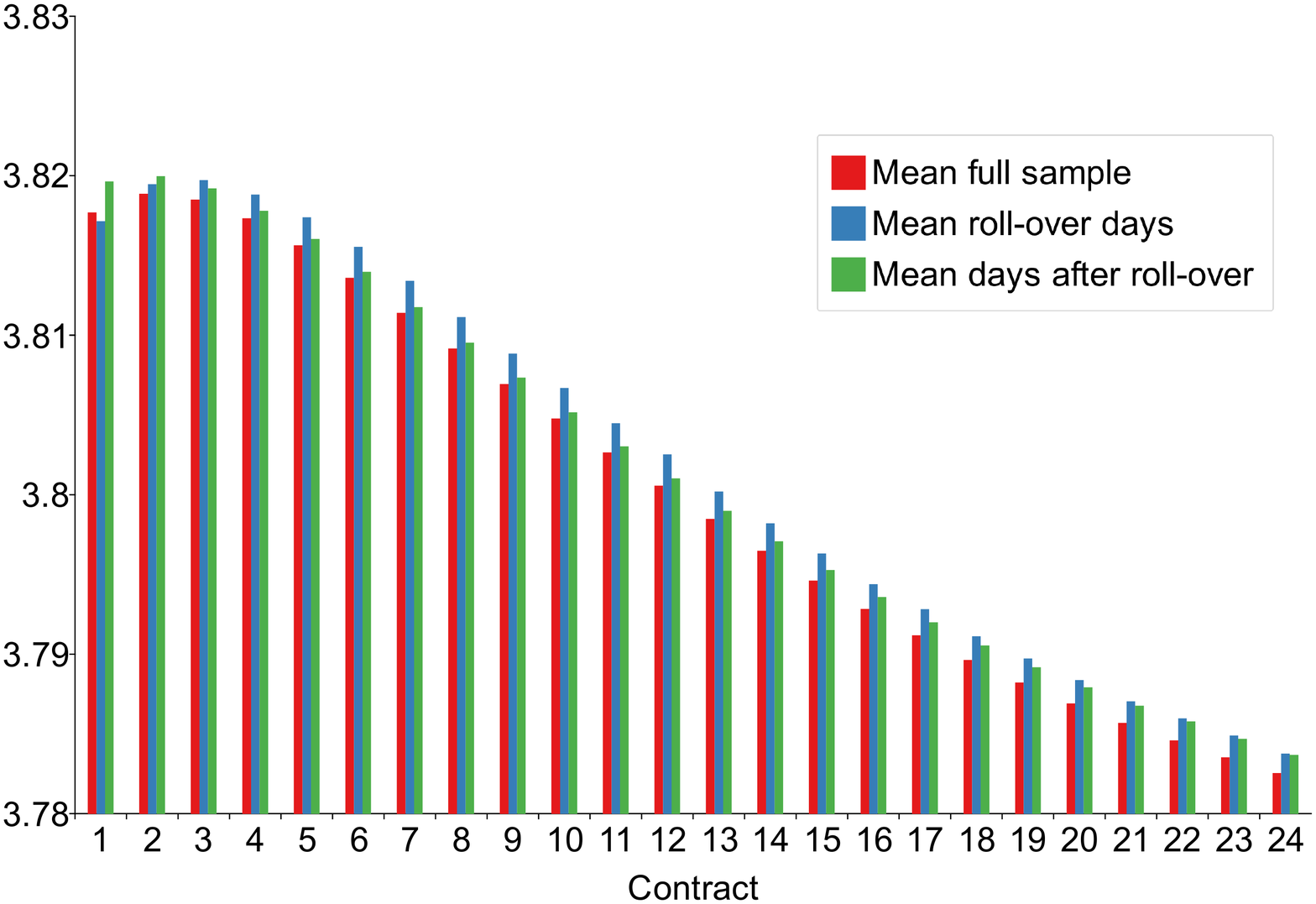} \\[-0.6cm]
\includegraphics[width=0.8\textwidth,angle=0]{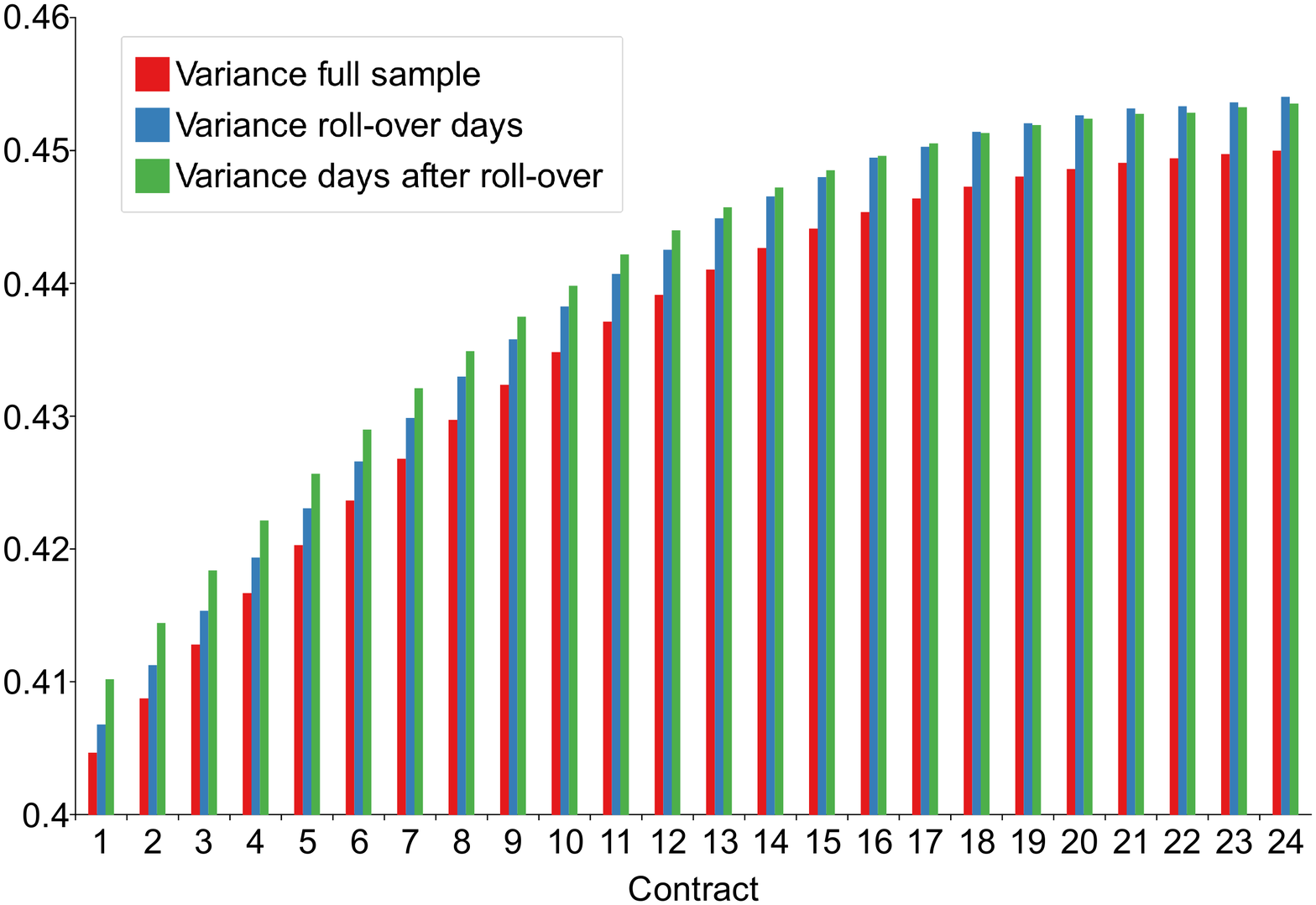}
\captionsetup{font=small}
\caption{Sample mean (upper panel) and variance (lower panel) of the log prices for the 24 monthly closest-to-delivery  crude oil future contracts computed for the full sample (red bar) together with the sample mean and variance of the log prices at the roll-over day (blue bar)  and one day after a roll-over (green bar).   The number of roll-over days is 244.  \label{fig:mean_var_data}}
\end{figure}
\end{center}

In Figure \ref{fig:data} we provide time-series plots of the log-prices  for the 24 contracts and in Figure
\ref{fig:mean_var_data}  the time average  of the term structure and its variation. The time-series plots in Figure \ref{fig:data}  reveal a strong co-movement of the prices for the 24 contracts suggesting  a common factor structure. We also observe substantial time variation in the term structure shape and  the price  volatility. The upper panel of Figure \ref{fig:mean_var_data} shows the time averages of the prices for the 24 contracts computed for the full sample. They reveal a slight concavity in the average term structure  with average  price levels which are monotonically decreasing in time to maturity for  maturities longer than 2 months.
As for the term structure of price variation, the   lower panel of Figure \ref{fig:mean_var_data} providing the sample variances  of the prices indicate that  the average price volatility is  increasing when moving from the short to the long end of the term structure.

As we are using time series of daily prices for perpetual contracts   obtained
by concatenating
contracts  with  maturities which differ by a month, the observed slopes in the term structure of the price level and the volatility can be expected to generate systematic discontinuities  in the  price processes when moving from a roll-over day to the next trading day (see also Ma et al., 1992).
In order to detect potential discontinuities of this sort
we computed the sample mean and variance of the prices observed at the
roll-over days ($t^*$)
and
for the days after the roll-over ($t^*+1$). The resulting statistics presented in Figure \ref{fig:mean_var_data}  show  that at the short end of the term structure the price variation at days after the roll-over is systematically higher relative to that at roll-over days with a difference which vanishes when moving to the long end. As  roll-overs increase time to maturity by a month this pattern is consistent with the upward-slope  observed for the average volatility term structure.
A similar systematic  pattern with differences at the short end which vanish at the long end we find for the average prices levels. Though not substantial in their order of magnitude at this aggregate level, those recurrent systematic jumps in the level and variation
generate an impact that is more visible from a closer inspection of the price movements around the 244 individual roll-over days we have in our sample period. Hence, dynamic term-structure models for time series of perpetual contracts should account not only for the observed slopes of the average term structure of the price level and its volatility but also
for those systematic discontinuities.

\section{Factor State-Space Models for the Term Structure}
\subsection{Factor structure}
The  econometric model we propose for the term structure of oil futures builds upon the  dynamic version of the NS model  introduced by Diebold and Li (2006) and Diebold et al.~(2006). While the (dynamic) NS model was initially designed to parsimoniously approximate yield curves, West (2012) and  Gr{\o}nberg und Lunde (2016) have recognized that it is also a useful approach for analyzing and forecasting  the term structure of futures on commodities.

Applied to crude oil future prices, the dynamic NS model has the form
\begin{align}\label{eq-3-Factor-TS-1}
y_{it}=\beta_{1t}+\beta_{2t}\left( \frac{1-e^{-\lambda_1 \tau_{it}}}{\lambda_1 \tau_{it}}\right)+
\beta_{3t}\left( \frac{1-e^{-\lambda_1 \tau_{it}}}{\lambda_1 \tau_{it}}-e^{-\lambda_1 \tau_{it}}\right)+\epsilon_{it}, \quad i=1,\ldots, N,
\end{align}
where $y_{it}$ denotes the period-$t$ log-price observed for  future contract $i$ with  maturity $\tau_{it}$, and $\epsilon_{it}$ is  an error term. The coefficients  $\beta_{t}=(\beta_{1t},\beta_{2t},\beta_{3t})'$ can be  interpreted as factors representing the level ($\beta_{1t}$), the slope ($\beta_{2t}$) and the curvature ($\beta_{3t}$) of the term-structure curve (Diebold and Li, 2006).
The factor loading for the level factor takes the value one, while  the loadings on the slope and curvature factor are functions of the contract-specific period-$t$ maturity $\tau_{it}$ with parameter $\lambda_1$. The parameter  $\lambda_1$ specifies the exponential decay rate  of the loadings along the term structure and determines how well the short end of the term structure is approximated. As it also determines for which maturity the hump-shaped loading of the curvature factor attains its maximum, $\lambda_1$ also controls the concavity of that loading.

As an extension of the  NS specification (\ref{eq-3-Factor-TS-1}) we consider the term-structure model introduced by Svensson (1994). In order to increase the range of potential shapes for the term-structure curve and improve the fit, Svensson (1994) propose to add to the NS model  a fourth term with an additional factor and decay parameter.
The dynamic version of the Svensson model is described as
\begin{align}\label{eq-4-Factor-TS-1}
y_{it}=  \beta_{1t}+\beta_{2t}\left( \frac{1-e^{-\lambda_1 \tau_{it}}}{\lambda_1 \tau_{it}}\right)&+\beta_{3t}\left( \frac{1-e^{-\lambda_1 \tau_{it}}}{\lambda_1 \tau_{it}}-e^{-\lambda_1 \tau_{it}}\right)\\
                &+ \beta_{4t}\left( \frac{1-e^{-\lambda_2 \tau_{it}}}{\lambda_2 \tau_{it}}-e^{-\lambda_2 \tau_{it}}\right)+\epsilon_{it}, \quad i=1,\ldots, N,\nonumber
\end{align}
with $\beta_{t}=(\beta_{1t},\beta_{2t},\beta_{3t},\beta_{4t})'$, where the fourth factor $\beta_{4t}$ can be interpreted as a second curvature factor.

\subsection{Factor dynamics and stochastic volatility}
Following Diebold et al.~(2006) we treat the factors $\beta_t$ as latent states so that we can take the NS and Svensson model  to formulate  dynamic state-space models (SSMs). (Applications of  SSMs designed for the analysis of yield curves  are found, e.g., in  Koopman et al., 2010, Hautch and Ou, 2012, and  Mesters et al., 2014.)
For this purpose, Equations (\ref{eq-3-Factor-TS-1}) and (\ref{eq-4-Factor-TS-1}) are written as
\begin{align}\label{eq-measurement-1}
y_t &= Z_t\beta_t+\epsilon_t,\qquad \epsilon_t\sim \mbox{i.i.d.} {\cal N}(0,\Sigma_y),\qquad t=1,\ldots,T,
\end{align}
with measurement vector $y_t=(y_{1t},\ldots,y_{Nt})'$ and measurement-error vector $\epsilon_t=(\epsilon_{1t},\ldots,\epsilon_{{N}t})'$, which is assumed to be  normally distributed.
$Z_t$ represents the $(N\times m)$-dimensional matrix of factor loadings, where $m$ is the number of factors. For the 4-factor Svensson model the $i$'th row of $Z_t$ is given by
\begin{align}\label{eq-loadings}
z_{it}' = \left(1\,,\,  \frac{1-e^{-\lambda_1 \tau_{it}}}{\lambda_1 \tau_{it}}\,,\,    \frac{1-e^{-\lambda_1 \tau_{it}}}{\lambda_1 \tau_{it}}-e^{-\lambda_1 \tau_{it}}\,,\,
\frac{1-e^{-\lambda_2 \tau_{it}}}{\lambda_2 \tau_{it}}-e^{-\lambda_2 \tau_{it}}\right),
\end{align}
and 3-factor NS version of $z_{it}$ obtains by excluding the last element.

For the $m$ latent factors $\beta_t$ we assume a Gaussian vector autoregressive (VAR) process of the form
\begin{align}\label{eq-beta-VAR}
\beta_t = \alpha + \Phi \beta_{t-1}  +\eta_t,\qquad \eta_t \sim {\cal N}(0,H_t^{-1}),
\end{align}
where $\alpha$ denotes an $(m\times 1)$ vector of intercepts,   $\Phi$  an $(m\times m)$ matrix with  slope coefficients, and  $H_t$ is an $(m\times m)$  precision matrix of the state innovations $\eta_t$.  The initial value $\beta_0$ is treated as a fixed parameter.
In order to account for  dynamic stochastic time-variation  in the volatility of the future prices $y_t$ we endow the factor innovations $\eta_t$ with a multivariate SV process
assuming that the volatility of future prices (given their maturities) is driven  by a dynamically evolving  uncertainty about  the level, the slope and the curvature of the term-structure curve (Hautsch and Ou, 2012; Gr{\o}nberg und Lunde, 2016). The model we assume for the volatility of $\eta_t$ is the multivariate Wishart SV process as introduced  by  Uhlig (1994, 1997) and further developed in Windle and Carvalho (2014).
It specifies the transition equation for the  precision matrix of $\eta_t$ as the following scaled singular Beta process:
\begin{align}\label{eq-transition-prec}
H_t = \frac{1}{\gamma}{H_{t-1}^{1/2}}'\Psi_t H_{t-1}^{1/2}, \qquad \Psi_t \sim {\cal B}_m\left(\frac{\nu}{2},\frac{1}{2}\right),\quad \nu >m-1,\;\gamma>0,
\end{align}
with parameters $\nu$ and $\gamma$ and  initial condition
\begin{align}\label{eq-transition-inits}
H_1|\Sigma_0 \sim  {\cal W}_m\left(\nu,\Sigma_0^{-1}/\gamma\right).
\end{align}
The matrix  $H_t^{1/2}$  is the upper Cholesky factor of $H_t$ and  ${\cal B}_m(\nu/2,1/2)$ denotes an $(m\times m)$-dimensional singular  beta distribution for the innovation matrix $\Psi_t$ with $\nu/2$ and $1/2$ degrees of freedom as defined in Uhlig (1994). ${\cal W}_m(\nu,\Sigma_0^{-1}/\gamma)$ represents an $(m\times m)$-dimensional Wishart distribution with $\nu$ degrees of freedom and scaling matrix $\Sigma_0^{-1}/\gamma$, where $\Sigma_0$ is an initial  $(m\times m)$ symmetric matrix.
The parameter $\nu$ determines (for a given value of $\gamma$) the conditional variation of $H_t$ given $H_{t-1}$: The  larger  $\nu$ is, the smaller the variation of the Beta innovation $\Psi_t$ with outcomes  which increasingly concentrate around the identity matrix, so that
$H_t=H_{t-1}/\gamma$ obtains as the limit for $\nu\to\infty$. The conditional expectation of $H_t$ is given by E$(H_t|H_{t-1})= H_{t-1}\nu/[\gamma(\nu+1)]$. Hence, the scaling parameter $\gamma$ together with $\nu$ controls the dynamics of the $H_t$ process, which becomes for $1/\gamma= (\nu+1)/\nu$ a martingale with E$(H_t|H_{t-1})= H_{t-1}$ .

Under the factor SSM model
for the daily prices of monthly contracts as defined by  Equations (\ref{eq-measurement-1})-(\ref{eq-transition-inits}), the mean and  variance of the prices  are given by
\begin{align}
\mbox{E}(y_{it})= z_{it}'\mbox{E}(\beta_t),\qquad \mbox{Var}(y_{it})= z_{it}'\mbox{Var}(\beta_t) z_{it}+\sigma_{y,i}^2,
\end{align}
respectively, where $\sigma_{y,i}^2$ denotes the variance of the contract-$i$ measurement error. Thus under the proposed  model it is not only the average price level which depends via the factor loadings $z_{it}$     on  the  period-$t$ time-to-maturity $\tau_{it}$ (see Equation \ref{eq-loadings}), but also the
volatility of the prices. Since the roll-overs used to obtain our price series $y_{it}$ increase the period-$t$ maturity by a month resulting in corresponding changes  in the factor loadings,
the model also predicts systematic jumps in the  price levels and volatility when moving from a roll-over day to the next day. Hence, the model is able to account not only for a trend in the price level and  volatility along the term structure but also for the systematic discontinuities in the level and volatility at the roll-over dates observed in the data (see Section 2).
This is illustrated in the online Appendix A5 where we provide plots of estimates for $\mbox{E}(y_{it})$ and $\mbox{Var}(y_{it})$.

\subsection{Parameter restrictions}
The factor SSM model as given by Equations (\ref{eq-measurement-1})-(\ref{eq-transition-inits}) contains the multivariate state processes ($\beta_t$, $H_t$) with initial conditions ($\beta_0$, $\Sigma_0$) and the parameters ($\lambda$, $\Sigma_y$, $\alpha$, $\Phi$, $\nu$, $\gamma$), where $\lambda=(\lambda_1,\lambda_2)'$.
In our application to the analysis of oil-future prices we deliberately impose the following restrictions on the parameters:  First, we use the restriction $\Phi=I_{m}$, where $I_{m}$ denotes the $m$-dimensional identity matrix, so that the VAR process for $\beta_t$ in Equation (\ref{eq-beta-VAR}) becomes a diagonal random walk.  That this is an empirically reasonable assumption, we have found in an initial explorative analysis  based on the $\beta_t$ factors extracted using period-by-period cross-sectional least squares.
Next, we assume that  the measurement errors are homoscedastic with a diagonal  covariance matrix $\Sigma_y=\sigma_y^2I_{N}$. The restriction of a diagonal form for $\Sigma_y$
is often used in order to reduce the number of parameters, which is critical in high-dimensional applications with a large number of assets $N$ (Diebold et al., 2006).
Finally, we impose on the parameters of the Wishart SV process in Equation (\ref{eq-transition-prec}) the restriction
\begin{align}\label{eq:wc-restriction}
\frac{1}{\gamma}=1+\frac{1}{\nu-m-1},
\end{align}
as proposed by Windle and Carvalho (2014). This constraint implies that $\gamma\to 1$ as $\nu\to\infty$, so that the covariance matrix of the factor innovations $\eta_t$ becomes time invariant with     $H_t^{-1}=\Sigma_0$ $\forall t$, in which case the factor SSM model in Equations (\ref{eq-measurement-1})-(\ref{eq-transition-inits})  reduces to a   standard linear-Gaussian SSM without stochastic volatility. The restriction (\ref{eq:wc-restriction}) also implies that the one-step-ahead forecast for the covariance matrix $H_{t+1}^{-1}$ based on $\beta_1,\ldots,\beta_t$ obtains as the following simple exponentially weighted moving average (EWMA) (Windle and Carvalho, 2014):
\begin{align}\label{eq:EWMA}
\mbox{E}(H_{t+1}^{-1}|\beta_{1:t})= (1-\gamma)\eta_t\eta_t'+\gamma \mbox{E}(H_{t}^{-1}|\beta_{1:t-1}),
\end{align}
where the notation  $A_{s:s'}$ is used  to denote  a  collection $\{A_s,\ldots,A_{s'}\}$.

Under this set of restrictions the list of parameters including  initial conditions, denoted by $\theta$, is given by $\theta=(\lambda,\sigma_y,\alpha,\nu,\beta_0, \Sigma_0)$.
With those restrictions our proposed model provides a very parsimonious yet flexible  framework for analyzing and forecasting the term structure of oil futures which is scalable in the number of assets $N$. A further important advantage  of our approach is that it is amenable to a computationally fast and easy to implement Bayesian posterior analysis  based on off-the-shelf MCMC procedures as discussed in the next section.

\section{Bayesian Posterior Analysis and Forecasting}
\subsection{MCMC algorithm}
For the Bayesian MCMC posterior analysis of the proposed non-linear non-Gaussian factor SSM  model
we use  a Gibbs approach to simulate from the joint posterior of the parameters and the states.
Using  $p(\cdot)$ to denote  a prior density, the joint posterior has the form,
\[
\pi(\beta_{0:T},H_{1:T},\lambda,\sigma_y^2,\alpha,\nu|y_{1:T})\propto f_\theta(y_{1:T}|\beta_{1:T}) f_\theta(\beta_{1:T}|H_{1:T}) f_\theta(H_{1:T})p(\beta_0)p(\Sigma_0)p(\lambda)p(\sigma_y^2)p(\alpha)p(\nu),
\]
where  $f_\theta(y_{1:T}|\beta_{1:T})$ and  $f_\theta(\beta_{1:T}|H_{1:T})$  are the conditional densities  for the  prices $y_{1:T}$ and the factors $\beta_{1:T}$, respectively, as defined by Equations (\ref{eq-measurement-1}) and (\ref{eq-beta-VAR}), and $f_\theta(H_{1:T})$ is the density of the precision matrices  $H_{1:T}$ according to Equations (\ref{eq-transition-prec}) and (\ref{eq-transition-inits}). In our application we select fairly uninformative priors except for the initial condition $\Sigma_0$, for which we use a degenerate prior with $\Sigma_0=0.1^2I_{m}$ (for details of prior selection, see the online Appendix A4).

A key feature of our proposed  Gibbs algorithm is that we exploit the  property of the factor SSM model  that it defines conditionally on $H_{1:T}$  a   SSM  for  $y_{1:T}$ with states given by $\beta_{1:T}$, and then
a SSM for $\beta_{1:T}$ with states $H_{1:T}$, such that both, $\beta_{1:T}$ as well as the $H_{1:T}$ can be directly simulated from their respective  exact conditional posterior distributions. Since the SSM for  $y_{1:T}$ given $H_{1:T}$     is   linear-Gaussian  the exact conditional posterior of $\beta_{1:T}$ obtains immediately from standard  Kalman-filter algebra.  Under the
SSM for $\beta_{1:T}$,  even though  not being  a  linear-Gaussian one,  the conditional posterior for $H_{1:T}$  is  also available  in a closed-form formula.
This results from the
fact that under the Wishart SV process (\ref{eq-beta-VAR})-(\ref{eq-transition-inits}) the filtering distribution for $H_t$ w.r.t.~$\beta_{1:t}$ obtains as a Wishart distribution which is  an associated conjugate distribution to
the Gaussian distribution  for the factor innovations  $\eta_t$ (Windle and Carvalho, 2014).
An additional feature of our algorithm is that we make use of collapsed Gibbs moves  to  sample the factors $\beta_{0:T}$ together with the factor-loading parameters $\lambda$ in one block and likewise the precisions $H_{1:T}$ together with their degrees-of-freedom parameter $\nu$ (Liu, 1994).  This blocking strategy  is known to increase the effectiveness in terms of mixing of the MCMC chain relative to the obvious approach to update $\beta_{0:T}$ and  $\lambda$ (and likewise $H_{1:T}$ and $\nu$) in two separate Gibbs blocks from their respective full conditional posteriors  (Chib et al., 2006).

The resulting Gibbs algorithm we propose  consists of the following updating steps:\\
{\sl 1.) Sampling of ($\beta_{0:T}$, $\lambda$):}  The collapsed Gibbs move for jointly simulating $\beta_{0:T}$ and  $\lambda$ from their joint conditional posterior consists of sampling $\lambda$ marginally of $\beta_{0:T}$ from
\begin{align}
\label{eq-post-lambda}\pi(\lambda| H_{1:T},\sigma_y^2,\alpha,\nu,y_{1:T})\propto
\Big[  \int f_\theta(y_{1:T}|\beta_{1:T})f_\theta(\beta_{1:T}|H_{1:T})p(\beta_0)d\beta_{0:T}\Big]p(\lambda)= f_\theta(y_{1:T}|H_{1:T})p(\lambda),
\end{align}
followed by simulating $\beta_{0:T}$ from its full conditional posterior,
\begin{align}
\label{eq-post-beta}
\pi(\beta_{0:T}|\lambda, H_{1:T},\sigma_y^2,\alpha,\nu,y_{1:T})&\propto f_\theta(y_{1:T}|\beta_{1:T})f_\theta(\beta_{1:T}|H_{1:T})p(\beta_0).
\end{align}
For simulating $\beta_{0:T}$ from its target density (\ref{eq-post-beta}), with $f_\theta(y_{1:T}|\beta_{1:T})$ and  $f_\theta(\beta_{1:T}|H_{1:T})$  defining
conditionally on  $H_{1:T}$ a standard linear-Gaussian SSM for $y_{1:T}$, we use the precision sampler of Chan and Jeliazkov (2009).
This  sampler exploits the sparsity of the precision matrix for $\beta_{0:T}$ under its Gaussian prior $f_\theta(\beta_{1:T}|H_{1:T})p(\beta_0)$ so that it is in  high-dimensional applications computationally  much faster than procedures based on standard Kalman filtering  such as
the algorithm of de Jong and Shephard (1995).
In order to  simulate  $\lambda$ from its target (\ref{eq-post-lambda}) we apply a Gaussian random-walk Metropolis-Hastings (RW-MH) algorithm (Chib and Greenberg, 1995). For its implementation,
the integrated likelihood $f_\theta(y_{1:T}|H_{1:T})$ for $y_{1:T}$ given $H_{1:T}$
is evaluated utilizing the method of Chan and Jeliazkov (2009) which is  based upon the same sparse Gaussian  algebra as it is  used for the  construction of their precision sampler for $\beta_{0:T}$.

{\sl 2.) Sampling of $(H_{1:T}$, $\nu$):} For sampling $H_{1:T}$ and $\nu$ from their joint conditional posterior, $\nu$ is simulated marginally of $H_{1:T}$ from
\begin{align}\label{eq-post-nu}
\pi(\nu|\beta_{0:T},\lambda,\sigma_y^2,\alpha,y_{1:T})&\propto
\Big[  \int f_\theta(\beta_{1:T}|H_{1:T})f_\theta(H_{1:T})dH_{1:T}\Big]p(\nu)=f_\theta(\beta_{1:T})p(\nu).
\end{align}
and the precision matrices $H_{1:T}$ from their full conditional posterior given by
\begin{align}
\label{eq-post-H}\pi(H_{1:T}|\nu, \beta_{0:T},\lambda,\sigma_y^2,\alpha,y_{1:T})&\propto
f_\theta(\beta_{1:T}|H_{1:T})f(H_{1:T}).
\end{align}
From Windle and Carvalho (2014, Propositions 1 and 2) it follows that the target density (\ref{eq-post-H}) under  the multivariate Wishart SSM
defined by Equations (\ref{eq-beta-VAR})-(\ref{eq-transition-inits}) and (\ref{eq:wc-restriction}) can be easily simulated by backward sampling from the following shifted rank-1 singular Wishart distribution:
\begin{align}
H_t|(H_{t+1},\eta_{1:t}) = \Big(\frac{\nu-m-1}{\nu-m}\Big)H_{t+1}+Z_{t+1},\qquad Z_{t+1}\sim {\cal W}_m(1,\Sigma_t^{-1}),
\end{align}
where the sequence of $\Sigma_t$'s obtains from forward filtering as $\Sigma_t=\eta_t\eta_t'+ [(\nu-m-1)/(\nu-m)]\Sigma_{t-1}$.
(For the definition of the singular Wishart ${\cal W}_m(1,\Sigma_t^{-1})$, see Uhlig, 1994.)
The forward filtering is initialized by $\Sigma_0$ as selected for the initial condition of $H_1$ as given in  Equation (\ref{eq-transition-inits}).
In order to simulate $\nu$ from the density (\ref{eq-post-nu}) we exploit that the integrated likelihood  $f_\theta(\beta_{1:T})$  for $\beta_{1:T}$ obtains as  the following  product of  conditional multivariate Student-$t$ densities for $\beta_t$ given $\beta_{0:t-1}$:
\begin{align}\label{eq:integr-eta-likelihood}
f_\theta(\beta_{1:T})&=\prod_{t=1}^T f_\theta(\beta_t|\beta_{0:t-1}),\quad \mbox{with}\\
f_\theta(\beta_t|\beta_{0:t-1})=\frac{\Gamma(\frac{\nu+1}{2})}{\pi^{m/2}\Gamma(\frac{\nu-m+1}{2})}&
\Big|\Big(\frac{\nu-m-1}{\nu-m}\Big)\Sigma_{t-1}\Big|^{-\frac{1}{2}}
\Big[1+\eta_t'\Big[\Big(\frac{\nu-m-1}{\nu-m}\Big) \Sigma_{t-1}\Big]^{-1} \eta_t\Big]^{-\frac{\nu+1}{2}}\nonumber,
\end{align}
where $|\cdot|$ is used to denote the determinant. (For the derivation of $f_\theta(\beta_{1:T})$, see online Appendix A1.)
This closed-form formula for the integrated likelihood  is used to simulate $\nu$ by a standard Gaussian RW-MH.

{\sl 3.) Sampling of $\alpha$ and $\sigma^2_y$:} Since we assume for $\alpha$ a conjugate Normal prior  and for  $\sigma_y^2$ a conjugate inverted Gamma prior, we can directly simulate from their full conditional posteriors $\pi(\alpha|\beta_{0:T},H_{1:T},\lambda,\sigma_y^2,\nu,y_{1:T})$ and $\pi(\sigma_y^2|\beta_{0:T},H_{1:T},\lambda,\sigma_y^2,\alpha,\nu,y_{1:T})$.

Our Gibbs algorithm cycles through Steps 1.) to 3.), and after dropping the draws from the first cycles as burn-in we use the draws from the next $S$ cycles for approximating the joint posterior. The posterior means of the model parameters and states used as point estimates are then approximated by the sample average over the respective $S$ Gibbs draws. We have implemented this Gibbs algorithm using MATLAB (version R2018b) on a 2016 iMac with a 3.1 GHz Intel Core i5 processor and 8GB of RAM. For the SSM with 4 factors and stochastic volatility  one Gibbs cycle through Steps 1.) to 3.) only takes 0.5 seconds.

For the restricted model without stochastic volatility this Gibbs algorithm is trivially adjusted by setting in Steps 1.) and 3.) the covariance matrix of the factor innovations to $H_{t}^{-1}=\Sigma_0$ $\forall t$, and replacing Step 2.) by a simulation step for $\Sigma_0$. Since we use for $\Sigma_0$ a conjugate inverted-Wishart prior it can be directly simulated from   its full conditional posterior.

\subsection{Model comparison}
In order to compare the 3-factor NS  with the 4-factor Svensson  version of the  proposed  SSM approach and to assess the importance of accounting for  stochastic volatility in the factors,  we use the deviance information criterion (DIC) (Spiegelhalter et al., 2002). The DIC measures the trade-off between model fit and model complexity and is given by
\begin{align}\label{eq-DIC-1}
\mbox{DIC}=-2\log f_{\hat \theta}(y_{1:T})+2 p_D,\quad p_D=-2\{\mbox{E}[\log f_\theta(y_{1:T})|y_{1:T}] - \log f_{\hat \theta}(y_{1:T})\},
\end{align}
with small values of the criterion preferred. $f_{\hat \theta}(y_{1:T})$ represents the likelihood function for  the observed data evaluated at the posterior estimates of the parameters $\hat \theta$ measuring the goodness of fit, and $p_D$ is the effective sample size which penalizes  rich parameterizations.  The term  $\mbox{E}[\log f_\theta(y_{1:T})|y_{1:T}]$  represents the mean of the log-likelihood function taken w.r.t.~the posterior distribution of $\theta$.

For the factor SSM model the data likelihood entering  the  DIC criterion  is given by the analytically intractable  high-dimensional integral
\begin{align}\label{eq-likelihood-data-1}
f_\theta(y_{1:T})=\int f_{\theta}(y_{1:T}|\beta_{1:T}) f_{\theta}(\beta_{1:T}|H_{1:T})f_\theta(H_{1:T})dH_{1:T}d\beta_{1:T}.
\end{align}
For its point-wise evaluation at given values of $\theta$,  we rely on
a Rao-Blackwellised  Sequential Monte-Carlo (SMC)  algorithm (Doucet and Johansen, 2009).
It exploits that according to Equations (\ref{eq-post-nu}) and (\ref{eq:integr-eta-likelihood}) the precision matrices $H_{1:T}$  can be integrated out  analytically (for details of its implementation, see  online Appendix A2).
Using this SMC for likelihood evaluations we can also estimate the posterior mean of the log-likelihood function in Equation (\ref{eq-DIC-1}) by the arithmetic mean over the Gibbs draws of the parameters $\{\theta^{(j)}\}$, i.e. $\hat{\mbox{E}}[\log f_\theta(y_{1:T})|y_{1:T}]=\frac{1}{S}\sum_{j=1}^S \log f_{\theta^{(j)}}(y_{1:T})$. (Note that for the restricted model without stochastic volatility the  likelihood according to Equations (\ref{eq-likelihood-data-1})   becomes the likelihood  of a linear-Gaussian SSM so that it can be evaluated by the method of  Chan and Jeliazkov (2009) mentioned above.)

\subsection{Forecasting}
In order to analyse the predictive performance of our proposed  model and to compare it with that of alternative forecasting models, the Gibbs algorithm  outlined in Section 4.2 can be used to compute out-of-sample point and density forecasts for the vector of prices $y_{t+1}$.

Under the  factor SSM model the one-step-ahead predictive density for $y_{t+1}$  is given by
\begin{align}\label{eq:predict_dens}
p(y_{t+1}|y_{1:t})&=
\int f_\theta(y_{t+1}|\beta_{t+1})f_\theta(\beta_{t+1}|\beta_t,H_{t+1})f_\theta(H_{t+1}|H_t)\\
&\qquad\qquad\qquad\qquad\qquad \times \pi(\beta_{1:t},H_{1:t},\theta|y_{1:t})d\beta_{1:t+1}dH_{1:t+1}d\theta, \nonumber
\end{align}
where $\pi(\beta_{1:t},H_{1:t},\theta|y_{1:t})$ is the   posterior
of the states and  parameters for the data up to period $t$. This predictive density  can be straightforwardly approximated using  direct MC integration
based on Gibbs simulation from the period-$t$ posterior $\pi(\beta_{1:t},H_{1:t},\theta|y_{1:t})$. (For details, see online Appendix A3.)
Note that such MC approximations to $p(y_{t+1}|y_{1:t})$
require to update in each period $t$ the  posterior $\pi(\beta_{1:t},H_{1:t},\theta|y_{1:t})$ by running period-by-period the full Gibbs sampler with the up-dated data set $y_{1:t}$. However, since our Gibbs algorithm is computationally very fast such updates are not very time consuming.
Based on those MC approximations for $p(y_{t+1}|y_{1:t})$ we can compute for an out-of-sample window $t=\{T+1,\ldots,T^*\}$ the log-predictive likelihood $\mbox{log-PL}=\sum_{t=T+1}^{T^*}\log  p(y_{t+1}|y_{1:t})$,
which we shall use for comparing the predictive performance of alternative versions of the factor SSM model.

For predictive performance comparisons with alternative non-Bayesian forecasting approaches we rely on one-step-ahead point and variance forecasts of the prices $y_{t+1}$ for a given value of the  parameters $\theta$, which obtain  as
\begin{small}
\begin{align}\label{eq-point-forecast}
\mbox{E}(y_{t+1}|y_{1:t},\theta)=Z_{t+1}\mbox{E}(\beta_{t+1}|y_{1:t},\theta),\quad
\mbox{Var}(y_{t+1}|y_{1:t},\theta)=Z_{t+1}\mbox{Var}(\beta_{t+1}|y_{1:t},\theta)Z_{t+1}'+\sigma_y^2I_N.
\end{align}
\end{small}
To approximate the  predictive moments $\mbox{E}(\beta_{t+1}|y_{1:t},\theta)$ and $\mbox{Var}(\beta_{t+1}|y_{1:t},\theta)$, we use direct MC-integration  based on Gibbs simulation from
the conditional period-$t$  posterior   $\pi(\beta_{1:t},H_{1:t}|y_{1:t},\theta)$  for a given value of $\theta$ (see online Appendix A3). The value for $\theta$ we use is its posterior mean estimate which we up-date for each time period $t$. The point and variance  forecasts in Equation (\ref{eq-point-forecast}) can also be used  for out-of-sample validation tests of the factor SSM model based on the predictive Pearson residuals, which are defined as
\begin{align}\label{eq-residuals}
\xi_{it+1} = [y_{it+1}-\mbox{E}(y_{it+1}|y_{1:t},\theta)]/\mbox{Var}(y_{it+1}|y_{1:t},\theta)^{1/2},\qquad i=1,\ldots, N.
\end{align}
If the predictive model is  valid, then $\xi_{it+1}$ has mean zero and unit variance, and $\xi_{it+1}$ as well as $\xi_{it+1}^2$ are serially uncorrelated.

\section{Empirical Results}
In this section, we apply our factor SSM model
to the daily log-prices of oil-future contracts described in Section 2. We consider two pairs of in-sample and out-of-sample periods (see Figure 1). The first pair consists of an in-sample period including all observations from Jan 2, 1996  up to Dec 31, 2007 with a sample size of $T=2998$, and an out-of-sample period from Jan 2, 2008 to Dec 31, 2008. Thus, this one-year out-of-sample window covers the 2008 financial crisis. The in-sample period of the second pair includes all observations from Jan 2, 1996 to  May 31, 2015 with a sample size $T=4865$, and the out-of-sample window covers the last year of our data  from Jun 1, 2015 to May 31, 2016.

In our application
we compare the following four specifications  w.r.t.~their empirical performance: The 3-factor NS model without stochastic volatility (3F) and with stochastic volatility  (3F-SV), and the  4-factor Svensson  model without (4F) and with stochastic volatility   (4F-SV).

\subsection{Estimation results}
For the Bayesian posterior analysis  we run the MCMC algorithm proposed in Section 4.1 for 11,000 Gibbs iterations, where the first 1,000 are discarded.
In order to evaluate the sampling efficiency of the MCMC procedure for estimating the parameters, we compute the effective sample size (ESS) of their posterior MCMC samples. It is defined as
$\mbox{ESS}=S[1+2\sum_{\ell=1}^L\hat\rho_\ell]^{-1}$, where $S$ is the MCMC posterior sample size, and $\sum_{\ell=1}^L\hat\rho_\ell$ represents the  $L$ monotone sample autocorrelations of the MCMC draws of a parameter (Geyer, 1992). The interpretation is that the $S$ MCMC draws  for a parameter lead to the same numerical precision as  a hypothetical iid sample from the posterior of size ESS, so that large ESS values are to be preferred.

The  posterior estimation  results for the four different factor SSM specifications obtained for the two in-sample periods are summarized in Table 1. The ESS values ranging from 202 to 10,000 indicate a fairly high sampling efficiency of the MCMC procedure with a sufficiently fast mixing rate for all the parameters.
We also find that the estimation results are fairly stable across the two sample periods with parameter estimates which are all reasonable. As indicated by the posterior standard deviations, the posterior uncertainty about the parameters is, apart from the VAR intercepts  $\alpha$, very low.
For the 3-factor models the estimates for the decay parameter of the loadings $\lambda_1$ range from 0.0054 to 0.0058. This is close to the value of 0.005 found by Gr{\o}nberg and Lunde (2016) using cross-section least squares for factor extraction from a 3-factor model applied to oil future prices from 2000 to 2010.
However, when moving from the 3-factor to the 4-factor models we find that the estimated $\lambda_1$ values significantly decrease and are substantially exceeded by the estimates for the second decay parameter $\lambda_2$. For example, under the 4F-SV specification fitted to the data of the second sample period, those estimates for $\lambda_1$ and $\lambda_2$ are 0.0036 and 0.0158, respectively. They imply that the loading of the first curvature factor $\beta_{3t}$ is maximized at a maturity of 500 days and that of the second curvature factor  $\beta_{4t}$ at 114 days. This indicates  that the two curvature factors under the 4-factor strucure are clearly distinguishable: The first curvature factor $\beta_{3t}$  is identified as a factor capturing a moderate concavity in the price curve, while the second one $\beta_{4t}$ appears to account for a strong concavity.
The estimates for the parameter of the Wishart SV process $\nu$ suggest that the factors exhibit significant  time-variation in their volatility with a fairly strong persistence.
According to restriction (\ref{eq:wc-restriction}), the estimates for $\nu$ imply that the EWMA smoothing parameter ($\gamma$) for the factors' conditional covariance matrix  in Equation (\ref{eq:EWMA}) range from 0.947 (3F-SV, second sample period)  to 0.958 (4F-SV, first sample period).  Such a range for the  EWMA parameter $\gamma$ is typical for  highly persistent time series.

As for the relative in-sample performance of the four specifications of the factor SSM, the DIC values reported in Table 1  show the following: In both sample periods, the 4-factor models yield a better trade-off between goodness of fit and parametric simplicity than their respective 3-factor counterparts, and models with stochastic volatility a better one than their counterparts without stochastic volatility. The model with the best trade-off is the 4F-SV. We also see that the gains in terms of DIC values obtained when including the fourth factor are larger than those achieved by adding stochastic volatility.
Thus, these  DIC results reveal strong evidence  against a 3-factor NS structure in favor of the 4-factor Svensson extension and
also corroborate our previous conclusion drawn from the estimates of the Wishart parameter $\nu$ that there is non-negligible dynamic time variation in the volatilities of the factors.

Figure 1 shows the time series plots of the posterior mean values for the factors $\beta_t$ obtained under the 4F-SV model. We observe that from the beginning of the 2000's until the 2008  financial crisis the level factor $\beta_{1t}$ is increasing, which reflects that prices for oil futures have steadily increased over that period. The slope factor $\beta_{2t}$  is most of the time positive implying a downward sloping term structure.  However, for the year  1999, after the crisis in 2008 and again in 2015, the $\beta_{2t}$ values indicate a  strong  upward slope in the  term structure.
Those results for the level and slope factor are fully in line with the 3-factor results of Gr{\o}nberg and Lunde (2016). For the two curvature factors $\beta_{3t}$ and $\beta_{4t}$ we find that they change their direction fairly frequently. However,  after the 2008 crisis the strong concavity factor $\beta_{4t}$ achieves its maximum indicating that the strong  upward slope in the term structure  is coupled with  a fairly pronounced hump-shaped curvature. For a plausibility check we  compared the posterior mean values of the factors plotted in Figure 1 with those of the extracted factors obtained from the cross-section least-squares approach (not presented here), which we have computed as described in Gr{\o}nberg and Lunde (2016).  It turned out that there are only small differences  between both types of factor estimates, which reflects that both estimation procedure
benefit from the fact that the observed prices are fairly informative about the factors.

Figure 3 displays for the 4F-SV  the smoothed estimates for the standard deviations and correlations of the factor innovations $\eta_t$ obtained from the posterior mean values of their covariance matrix $H_{t}^{-1}$. It also shows
non-parametric estimates for those standard deviations and correlations based on the factors extracted by  cross-section least squares. They  are obtained from the scaled realized covariance matrices based on a two-sided rolling window and are computed as $\mbox{RC}_t=\sum_{\ell=t-L}^{t+L}(\beta_\ell^e-\beta_{\ell-1}^e)(\beta_\ell^e-\beta_{\ell-1}^e)'/(2L+1)$, where $\beta_\ell^e$ denotes the vector of extracted factors and $L$ is the window size which we set equal to 6. (For details of realized covariances, see Barndorff-Nielsen and Shephard, 2004.)   From the time series plots of
the non-parametric estimates
it is evident that there is substantial time variation in the volatilities of the factors as well as in their correlations. We also see that the smoothed estimates of the volatilities and correlations obtained under the 4F-SV track the temporal evolution in the non-parametric estimates quite well. This illustrates  the  flexibility of the  parsimonious Wishart SV model in capturing the dynamics of multivariate volatility processes.

\subsection{Out-of-sample forecasting results}
We now analyse the
forecasting performance of the factor SSM.
For this
we rely on statistical as well as economic forecast evaluations.  For the former we use the predictive densities, diagnostic checks on the predictive residuals and the accuracy of point forecasts for the prices, while for the latter  we analyse the ability to predict the Value at Risk (VaR) for portfolios constructed from contracts with different delivery dates.
\subsubsection{Statistical forecast evaluations}
Table 2 reports the log-predictive likelihood values
for the four specifications of the  factor SSM  in the two out-of-sample windows, and Figure 4 shows the time-series plots of the  period-wise accumulated log-predictive densities for the 3F, 3F-SV and 4F model in terms of their difference relative to those of the 4F-SV.
The log-predictive likelihood values in Table 2  reveal that in both out-of-sample periods the 4F-SV  yields the best predictive fit. However,  
adding the fourth factor appears to be more important for the out-of-sample fit than including stochastic volatility. Thus, these results mirror our previous ones on the relative in-sample performance of the factor SSM specifications.
From  the accumulated 
log-predictive densities in Figure 4 we see that the gains in the out-of-sample fit obtained by accounting for stochastic volatility  are larger for the 2008 crisis period  than for the 2015-2016 period, while it is for the 2015-2016 period where the gains achieved by including the fourth factor are  the larger ones.

In Tables 2 and 3 we summarize the results of the out-of-sample diagnostic checks based on the predictive Pearson residuals $\xi_{it+1}$.
The means of these residuals reported in Table 2 show that all factor SSM models tend to slightly  overpredict the price levels. We also find that for the models without
stochastic volatility the standard deviations substantially  exceed the benchmark value of 1 under a correctly specified model, especially  for the 2008 crisis period. Hence, there is much more variation in the prices than those models predict. In contrast, the standard deviation we observe for the models including stochastic volatility are quite close to that benchmark value.  Table 3 provides the $p$-values of the Ljung-Box test with 10 lags for the
residuals $\xi_{it+1}$ and their squares $\xi_{it+1}^2$.
The $p$-values for $\xi_{it+1}^2$ show that
the specifications devoid of stochastic volatility  are not able to capture  the observed dynamics in the volatility of the prices.
Moreover,  in the 2008 period,  we find for those specifications significant correlation in $\xi_{it+1}$ for the contracts at the short  end of the term structure suggesting that they are also not able to fully account for the  dynamics in the corresponding price levels. The models with stochastic volatility satisfactorily fit the observed serial correlation in the price levels for both out-of-sample periods and also successfully capture the dynamics in the volatility during the 2015-2018 period. However, in the 2008 crisis  period, they appear to have problems to fully account for the observed volatility dynamics as it is indicated by  the significant correlation in $\xi_{it+1}^2$ for the contracts at  medium and long maturities.

In order to assess the accuracy of the one-day-ahead point forecasts for the prices of the 24 future contracts (see Equation \ref{eq-point-forecast}) we use the root mean squared forecast error (RMSFE). The results are summarized in Table 4 which reports the average RMSFE across all contracts and the average RMSFE for three groups of contracts  representing the short end, the medium part and the long end of the term structure. As benchmark values  we also report the corresponding RMSFE values for point forecasts obtained by using independent random walks, one for each of the 24 contracts. The results show that in both out-of-sample windows and for all parts of the term structure the accuracy of the four specifications of the factor-SSM is virtually the same. We also observe that the random-walk forecasts  are in all cases  more accurate than those of the factor-SSM specifications. However, the differences are immaterial. This is a positive outcome for the factor-SSM approach  since according to the efficient market hypothesis (Fama, 1970) the random-walk can be considered to represent a competitive and hard to beat benchmark for predicting  speculative prices  like those for oil-futures.

\subsubsection{Value-at-risk forecasts}
For our VaR application we consider two different portfolios constructed from the 24 contracts. The first one is an equally-weighted portfolio  and the second one a `bull-spread' portfolio consisting of a long position in the first contract (one month ahead) and a short position in the eighth contract (eight months ahead). The portfolio return realized at period $t+1$ is $r_{t+1}^p=\omega'(y_{t+1}-y_t)$, where $\omega$ denotes the vector of portfolio weights. Given the one-day-ahead forecasts for the contract prices $\mbox{E}(y_{t+1}|y_{1:t},\theta)$ and their covariance matrix $\mbox{Var}(y_{t+1}|y_{1:t},\theta)$ (see Equation \ref{eq-point-forecast}), the predicted portfolio return and its variance are
$\mu_{t+1|t}=\omega'[\mbox{E}(y_{t+1}|y_{1:t},\theta)-y_t]$ and $\sigma_{t+1|t}^2=\omega'\mbox{Var}(y_{t+1}|y_{1:t},\theta)\omega$, respectively. Using those predictive moments the one-day-ahead forecast   for the portfolio VaR at level $\alpha^*$ obtains as
\begin{align}\label{eq-VaR}
\mbox{VaR}_{t+1|t}(\alpha^*)=\mu_{t+1|t}+\sigma_{t+1|t}F^{-1}_{t+1}(\alpha^*),
\end{align}
where $F_{t+1}$ is the cumulative distribution function (cdf) of  the conditional distribution for the standardized returns $(r_{t+1}^p-\mu_{t+1|t})/\sigma_{t+1|t}$ given $y_{1:t}$. Common practice for calculating the VaR forecast is to use in Equation (\ref{eq-VaR}) (MC-)estimates of the predictive moments $\mu_{t+1|t}$ and $\sigma_{t+1|t}$. For the factor SSM  such estimates can be obtained from our MC approximations of the point and variance price forecasts $\mbox{E}(y_{t+1}|y_{1:t},\theta)$ and $\mbox{Var}(y_{t+1}|y_{1:t},\theta)$.
However, under the factor SSM the cdf $F_{t+1}$
is unknown. Hence, we take a direct MC approach to compute the  forecast $\mbox{VaR}_{t+1|t}(\alpha^*)$.
Using our Gibbs algorithm we  simulate     from the predictive density of the factors $f_\theta(\beta_{t+1}|y_{1:t})$, use the resulting $\beta_{t+1}$-draws to produce simulated portfolio  returns  from their predictive density $f_\theta(r_{t+1}^p|y_{1:t})$, and then calculate the $\alpha^*$-quantile from the empirical distribution of the simulated portfolio returns (see the Appendix A3).

Following Gr{\o}nborg and Lunde (2016), we compare the factor SSM  VaR forecasts  to those obtained from using  a  homoscedastic Gaussian VAR for the factors extracted  by cross-section least squares ($\beta_t^e$) given by $\beta_t^e= \mu+\Omega\beta_{t-1}^e+\xi_t$, with  $\xi_t \sim {\cal N}(0,\Sigma_\xi)$.
As for the factor SSM, we consider a three factor (VAR-3F) and a four factor version (VAR-4F).
The predicted portfolio return $\mu_{t+1|t}$ and variance $\sigma_{t+1|t}^2$ in the VaR equation (\ref{eq-VaR}), which obtains from combining  this VAR with with the measurement equation (\ref{eq-measurement-1}) for the prices, are
$\mu_{t+1|t}= \omega'[Z_{t+1}(\mu+\Omega\beta_{t}^e) -y_t]$ and $\sigma_{t+1|t}^2=\omega'[Z_{t+1}\Sigma_\xi Z_{t+1}' +\Sigma_y]\omega$, and $F_{t+1}$ is the cdf of a ${\cal N}(0,1)$ distribution.
Just as for the point forecasts  analyzed in the previous section, the VaR forecasts of the factor SSM and VAR model are computed by updating the parameter estimates at a daily frequency.

For assessing the accuracy  of the predicted VaR we  test for unconditional and  conditional coverage based on the `hit-indicator' variable $I_{t+1}=\mathds{1} [r_{t+1}^p\leq \mbox{VaR}_{t+1|t}(\alpha^*)]$, signaling that the realized portfolio return is lower or equal than the predicted VaR. Under a valid predictive model for the VaR the sequence of $I_{t+1}$'s is serially independent  and have the correct coverage such that the expectation of $I_{t+1}$ is equal to the nominal level $\alpha^*$. The test for unconditional coverage used is the likelihood-ratio (LR) test of Kupiec (1995) which compares the time average of  $I_{t+1}$ (i.e.~the hit rate) with the nominal value $\alpha^*$. For checking the independence hypothesis we apply the LR test proposed by Christoffersen (1998) testing the serial independence of $I_{t+1}$ against the alternative of  first-order Markov dependence. In order to test for  conditional coverage we  rely on the  LR test of Christoffersen (1998) which consists of a  combination of the LR tests for unconditional coverage and independence.

Table 5 provides the hit rate and  $p$-value for the tests of unconditional coverage, independence and conditional coverage for the 1\%, 5\% and 10\% VaR predicted for the equally-weighted portfolio. The corresponding figures for the  bull-spread portfolio are given in Table 6. The results reinforce our earlier ones on the predictive fit of the factor SSM. The best performing factor SSM for both portfolios and both out-of-sample periods is the 4F-SV. For the equally-weighted portfolio it predicts (according to the realized hit rates) convincingly the VaR at all considered $\alpha^*$ levels and passes  the tests for unconditional and conditional coverage as well as the  independence test at the 1\% significance level. However, for the bull-spread portfolio we find that the 4F-SV forecasts for the 10\% VaR  fail to pass in both periods the conditional coverage test. For the 2008 period this appears to be due to serial dependence in the predicted hit events, while for the 2015-2016 period the reason for failing is that the forecasts tend to be too conservative. The results also show that the relatively  good performance of the 4F-SV  in predicting   the equally-weighted portfolio
is mainly attributed to its stochastic volatility component. Without such a component the  VaR forecasts of the factor SSM  suffer from systematic  underpredictions, especially in 2008 crisis period. For the bull-spread portfolio it is mainly the fourth factor which appears to be critical for reliable VaR forecasts. Without the fourth factor the factor SSM predicts hit events which are in both periods significantly  serially dependent, and for the 2015-2016 window the resulting hit rate also substantially exceeds the nominal VaR level $\alpha^*$. As for the comparison of the factor SSM  with the VAR for the extracted factors, we observe that for the equally-weighted portfolio the best performing factor SSM  (4F-SV) compares favorably  to the VAR-3F and VAR-4F, especially during the 2008  period and is on par with the VAR-3F for the 2015-2016 period. In case of the bull-spread portfolio,
the performance of the 4F-SV during the 2008 period is not worse than that of the two VAR models,  but it is outperformed in the 2015-2016 forecast window by the VAR-4F.

\section{Conclusions}
From our findings we can draw three kind of conclusions. With respect to modelling   the term structure of contracts on commodities, we have illustrated  that our parsimonious latent factor state-space approach
combining the dynamic 3-factor Nelson-Siegel model and its 4-factor Sevensson extension  with a multivariate Wishart stochastic volatility process
is able to represent non-trivial joint dynamics in  multiple time series  of contract prices and provides a useful tool for  predicting the level and volatility of contract prices.

In regard to our application to 24 monthly future contracts on crude-oil  itself, we find robust  evidence against
the 3-factor Nelson-Siegel  structure  in favour of the 4-factor model \`{a} la Svensson, which adds to the Nelson-Siegel factors for the level, slope and  curvature  an additional curvature factor.
The two curvature factors appear to be clearly empirically distinguishable,  whereby one factor is identified to represent a moderate and the other one a strong concavity in the term structure.
We also find significant  time variation in the volatility and correlations of the four factors which is for the most part well captured by the Wishart  volatility process.
Both, the fourth factor as well as the stochastic volatility component proves to be important not only for the in-sample performance
but also for the out-of-sample predictive fit  and the ability   to predict the value-at-risk of portfolios of contracts with different delivery dates.

From the viewpoint of computational statistics, we demonstrate that our proposed high-dimensional factor model, though belonging to the class of non-linear non-Gaussian state-space models, is amenable to a computationally fast and easy to implement Bayesian posterior analysis based on tuned MCMC procedures. The specific MCMC algorithm we propose  exploits that under our model the conditional posterior distributions of all  latent states are available  in closed-form formulae. As such it can also be used to conveniently compute useful statistics for an out-of-sample analysis such as predictive densities, point and  variance forecasts and out-of-sample validation statistics.

Last but not least, the findings from the application to crude-oil future contracts lead us to believe that our proposed approach is also viable for the analysis of future contracts on other  energy commodities (such as natural gas and  heating oil) as well as  other groups of commodities (such as metals and agriculturals). In contrast to crude-oil,  future prices of many energy and agricultural commodities display distinct seasonal fluctuations  in which case our proposed factor SSM model needs to be extended. Initial experimentations with data on cotton futures show that extending the VAR equation  for the factors by  seasonal components successfully isolates meaningful seasonal cycles over the year.

\section*{References}
\begin{small}
\begin{description}
\item[]Barndorff-Nielsen, O.E, and Shephard, N., 2004. Econometric analysis of realized covariation: high frequency based covariance, regression, and correlation in financial economics. Econometrica 72, 885-925.\\[-1cm]
\item[]Barun\'{\i}k, J., and  Malinsk\'{a}, B., 2016. Forecasting the term structure of crude oil futures prices with neural networks. Applied Energy 164, 366-379.\\[-1cm]
\item[]Black, F., 1976. The pricing of commodity contracts. Journal of Financial Economics 3, 167-179.\\[-1cm]
\item[]Brennan, M. J., 1991. The price of convenience and the valuation of commodity contingent claims. In: Lund, D., and Oeksendal, B.~(eds), Stochastic Models and Option Values. Elsevier North Holland, 33-71.\\[-1cm]
\item[]Chan, J.C.C., and Jeliazkov, I., 2009. Efficient simulation and integrated likelihood estimation in state space models. International Journal of Mathematical Modelling and Numerical Optimisation 1, 101-120.\\[-1cm]
\item[]Chib, S., and Greenberg, E., 1995. Understanding the Metropolis-Hastings algorithm. The American Statistician 49, 327-335.\\[-1cm]
\item[]Chib, S., Nardari, F., and  Shephard, N., 2006. Analysis of high dimensional multivariate stochastic volatility models. Journal of Econometrics 134,  341-371.\\[-1cm]
\item[]Christoffersen, P., 1998. Evaluating interval forecasts. International Economic Review 39, 841-862.\\[-1cm]
\item[]De Jong, P., and Shephard, N., 1995. The simulation smoother for time series models. Biometrika 82, 339-350.\\[-1cm]
\item[]Diebold, F.X., and  Li,  C., 2006.  Forecasting the term structure of government bond yields.  Journal of Econometrics 130,  337-364.\\[-1cm]
\item[]Diebold, F.X., Rudebusch, G.D., and Aruoba, S., 2006. The macroeconomy and the yield curve. Journal of Econometrics  131, 309-338.\\[-1cm]
\item[]Doucet, A., and Johansen, A.M. (2009). A tutorial on particle filtering and smoothing: Fifteen years later.
       In: Crisan, D., Rozovskii, B. (eds), The Oxford Handbook of Nonlinear Filtering. Oxford University Press, 656-704.\\[-1cm]
\item[]Etienne, X.L., and Mattos F., 2016. The information content in the term structure of commodity prices. Proceedings of the NCCC-134 Conference on Applied Commodity Price Analysis, Forecasting, and Market Risk Management. St.~Louis, MO.\\[-1cm]
\item[]Fama, E., 1970. Efficient capital markets: A review of theory and empirical work. Journal of Finance 25, 383-417.\\[-1cm]
\item[]Geyer, C.J., 1992. Practical Markov Chain Monte Carlo. Statistical Science 7, 473-483.\\[-1cm]
\item[]Gr{\o}nborg, N.S., and Lunde, A., 2016. Analyzing oil futures with a dynamic Nelson-Siegel model. The Journal of Futures Markets  36, 153-173.\\[-1cm]
\item[]Hautsch, N., and Ou, Y., 2012. Analyzing interest rate risk: Stochastic volatility in the term structure of government bond yields. Journal of Banking \& Finance 36, 2988-3007.\\[-1cm]
\item[]Karstanje, D., van der Wel, M., and van Dijk, D.J.C., 2017. Common factors in commodity futures curves.  Working paper (available at SSRN: https://ssrn.com/abstract=2558014).\\[-1cm]
\item[]Koopman, S.J., Mallee, M.I.P., and van der Wel, M., 2010. Analyzing the term structure of interest rates using the dynamic Nelson-Siegel model with time-varying parameters. Journal of Business and Economic Statistics  28, 329-343.\\[-1cm]
\item[]Kupiec, P.H., 1995. Techniques for verifying the accuracy of risk management models. Journal of Derivatives 3, 73-84.\\[-1cm]
\item[]Liu, J.S.,  1994. The collapsed Gibbs sampler in Bayesian computations with applications to a gene regulation problem. Journal of the American Statistical Association 89, 958-966.\\[-1cm]
\item[]Ma, C.K., Mercer, J.M., and Walker, M.A., 1992. Rolling over futures contracts: A note. The  Journal of Future Markets 12, 203-217.\\[-1cm]
\item[]Mesters, G., Schwaab, B., and Koopman, S.J., 2014. A dynamic yield curve model with stochastic volatility and non-Gaussian interactions: An empirical study of non-standard monetary policy in the Euro area. Tinbergen Institute Discussion Paper 14-071/III.
\item[]Nelson, C.R., and Siegel, A.F., 1987. Parsimonious modeling of yield curves. The Journal of Business  60, 473-489.\\[-1cm]
\item[]Oglend, A., and Kleppe, T.S., 2019. Can limits-to-arbitrage from bounded storage improve commodity term-structure modeling? Journal of Futures Markets 39, 865-889.\\[-1cm]
\item[]Spiegelhalter, D.J., Best, N.G., Carlin, B.P., and van der Linde, A., 2002. Bayesian measures of model complexity and fit. Journal of the Royal Statistical Society - Series B  64, 583-639.\\[-1cm]
\item[]Svensson,  L.E.O., 1994. Estimating and interpreting forward interest rates: Sweden 1992-1994. NBER Working Paper Series No. 4871.\\[-1cm]
\item[]Uhlig,  H., 1994. On singular Wishart and singular multivariate Beta distributions. The Annals of Statistics 22, 395-405.\\[-1cm]
\item[]Uhlig,  H., 1997. Bayesian vector autoregressions with stochastic volatility. Econometrica 65, 59-73.\\[-1cm]
\item[]West, J., 2012. Long-dated agricultural futures price estimates using the seasonal Nelson-Siegel model. International Journal of Business and Management  7, 78-93.\\[-1cm]
\item[]Windle, J., and Carvalho, C.M., 2014. A tractable state-space model for symmetric positive-definite matrices. Bayesian Analysis 9, 759-792.\\[-1cm]
\end{description}
\end{small}



%
%

\pagebreak
\begin{table}[h!]\centering
\footnotesize{
\begin{tabular}{lccccccccc}
\multicolumn{10}{c}{\mbox{\small{\sl Table 1. MCMC posterior analysis of the factor SSM model}}}\\[0.1cm]\hline\\[-0.3cm]
  &\multicolumn{1}{c}{3F}  & \multicolumn{1}{c}{3F-SV} &\multicolumn{1}{c}{4F} &\multicolumn{1}{c}{4F-SV} &&\multicolumn{1}{c}{3F}  & \multicolumn{1}{c}{3F-SV} &\multicolumn{1}{c}{4F} &\multicolumn{1}{c}{4F-SV} \\
\hline\\

&\multicolumn{4}{c}{Jan 2, 1996 - Dec 31, 2007}&&\multicolumn{4}{c}{Jan 2, 1996 - May 29, 2015}\\
\cline{2-5}\cline{7-10}\\
$\alpha_1\times 100$
                       &   0.051     &   0.039   & 0.047   & 0.028 &&    0.027     &   0.020     & 0.027     & 0.015  \\
                       &   (0.029)   &   (0.018)   & (0.030)   & (0.019) &&    (0.022)     &   (0.013)     & (0.022)     & (0.012) \\
                       &\multicolumn{1}{c}{[9,907]}
                                                      &\multicolumn{1}{c}{[9,435]}
                                                                   &\multicolumn{1}{c}{[9,821]}
                                                                              & \multicolumn{1}{c}{[5,653]}
                                                                             && \multicolumn{1}{c}{[9,572]}
                                                                                  &\multicolumn{1}{c}{[7,819]}
                                                                                             &\multicolumn{1}{c}{[9,440]}
                                                                                                 & \multicolumn{1}{c}{[5,616]}\\[0.2cm]
$\alpha_2\times 100$
                      &   0.002     &   -0.013    & 0.005   & -0.017 && -0.005     &   0.003     &-0.005    & -0.002  \\
                      &   (0.039)     &    (0.026)    & (0.041)   &  (0.026) &&  (0.028)     &   (0.016)     & (0.030)    &  (0.016)  \\
                             &\multicolumn{1}{c}{[10,000]}
                                                      &\multicolumn{1}{c}{[8,732]}
                                                                   &\multicolumn{1}{c}{[9,967]}
                                                                              & \multicolumn{1}{c}{[6,310]}
                                                                                  && \multicolumn{1}{c}{[10,000]}
                                                      &\multicolumn{1}{c}{[8,968]}
                                                                   &\multicolumn{1}{c}{[9,868]}
                                                                              & \multicolumn{1}{c}{[5,680]} \\[0.2cm]

$\alpha_3\times 100$
                     &   0.010     &   0.005     & 0.014    & 0.030 &&   0.002     &   0.008     & 0.005  & 0.017 \\
                     &   (0.041)     &   (0.022)     & (0.068)    & (0.034) &&   (0.029)     &   (0.016)     & (0.038)  & (0.018) \\
                           &\multicolumn{1}{c}{[9,871]}
                                                      &\multicolumn{1}{c}{[7,436]}
                                                                   &\multicolumn{1}{c}{[9,920]}
                                                                              & \multicolumn{1}{c}{[5,528]}
                                                                                 &&
                                                                                 \multicolumn{1}{c}{[8,888]}
                                                      &\multicolumn{1}{c}{[7,398]}
                                                                   &\multicolumn{1}{c}{[10,000]}
                                                                              & \multicolumn{1}{c}{[5,917]}\\[0.2cm]

$\alpha_4\times 100$
                     &            &              & 0.005    & 0.007  &&            &            & 0.003    & 0.003 \\
                       &            &            & (0.047)    & (0.024)  &&            &            & (0.032)    & (0.011)\\
                             &\multicolumn{1}{c}{}
                                                      &\multicolumn{1}{c}{}
                                                                   &\multicolumn{1}{c}{[10,000]}
                                                                              & \multicolumn{1}{c}{[4,975]}
                                                                                 &&
                                                                                 \multicolumn{1}{c}{}
                                                      &\multicolumn{1}{c}{}
                                                                   &\multicolumn{1}{c}{[9,593]}
                                                                              & \multicolumn{1}{c}{[4,066]} \\[0.2cm]

$\lambda_1\times 100$
                   &   0.576     &   0.582     & 0.283     & 0.309  &&  0.541    &   0.545      & 0.335    & 0.359 \\
                   &   (0.003)     &   (0.003)     & (0.005)     & (0.006)  &&  (0.002)    &   (0.002)      & (0.004)    & (0.004)  \\
                              &\multicolumn{1}{c}{[2,193]}
                                                      &\multicolumn{1}{c}{[1,934]}
                                                                   &\multicolumn{1}{c}{[528]}
                                                                              & \multicolumn{1}{c}{[202]}
                                                                                    &&
                                                                                    \multicolumn{1}{c}{[2,286]}
                                                      &\multicolumn{1}{c}{[1,729]}
                                                                   &\multicolumn{1}{c}{[470]}
                                                                              & \multicolumn{1}{c}{[373]}    \\[0.2cm]
$\lambda_2\times 100$
                       &            &            & 1.544    & 1.545    &&           &            &  1.570   & 1.576  \\
                        &            &            &  (0.015)    &  (0.015)  &&           &            &   (0.010)   &  (0.011)\\
                              &\multicolumn{1}{c}{}
                                                      &\multicolumn{1}{c}{}
                                                                   &\multicolumn{1}{c}{[1,129]}
                                                                              & \multicolumn{1}{c}{[996]}
                                                                                   &&
                                                                                   \multicolumn{1}{c}{}
                                                      &\multicolumn{1}{c}{}
                                                                   &\multicolumn{1}{c}{[1,188]}
                                                                              & \multicolumn{1}{c}{[934]}  \\[0.2cm]
$\sigma_y\times 100$
                      &   0.429     &   0.429     & 0.399     & 0.398  &&  0.351      &   0.351      & 0.316     & 0.316 \\
                       &   (0.001)     &   (0.001)     & (0.001)     & (0.001)  &&  (0.0005)   &     (0.0005)   &   (0.0005)   &  (0.0005)\\
                        &\multicolumn{1}{c}{[8,690]}
                                                      &\multicolumn{1}{c}{[9,126]}
                                                                   &\multicolumn{1}{c}{[8,854]}
                                                                              & \multicolumn{1}{c}{[8,949]}
                                                                                    &&
                                                                                    \multicolumn{1}{c}{[8,942]}
                                                      &\multicolumn{1}{c}{[9,074]}
                                                                   &\multicolumn{1}{c}{[8,295]}
                                                                              & \multicolumn{1}{c}{[9,200]} \\[0.2cm]

$\nu$                  &            &   25.86    &          & 27.72  &&            &   21.75    &          & 23.97 \\
                        &            &    (1.48)    &          & (1.27)   &&            &     (0.76)    &          &   (0.70)  \\
                            &\multicolumn{1}{c}{}
                                                      &\multicolumn{1}{c}{[1,034]}
                                                                   &\multicolumn{1}{c}{}
                                                                              & \multicolumn{1}{c}{[784]}
                                                                                 &&
                                                                                 \multicolumn{1}{c}{}
                                                      &\multicolumn{1}{c}{[1,192]}
                                                                   &\multicolumn{1}{c}{}
                                                                              & \multicolumn{1}{c}{[825]}\\[0.8cm]
\multicolumn{1}{l}{$\log p_{\hat\theta}(y_{1:T})$}
                                        &\multicolumn{1}{c}{294,390}
                                                      &\multicolumn{1}{c}{295,235}
                                                                   &\multicolumn{1}{c}{307,270}
                                                                              & \multicolumn{1}{c}{308,366}
                                                                                &&
                                                                                \multicolumn{1}{c}{492,211}
                                                      &\multicolumn{1}{c}{494,170}
                                                                   &\multicolumn{1}{c}{520,833}
                                                                              & \multicolumn{1}{c}{523,231}\\[0.2cm]
DIC
                                        &\multicolumn{1}{c}{-588,835}
                                                      &\multicolumn{1}{c}{-590,456}
                                                                   &\multicolumn{1}{c}{-614,532}
                                                                              & \multicolumn{1}{c}{-616,742} &&  \multicolumn{1}{c}{-984,443}
                                                                                                                        &\multicolumn{1}{c}{-988,335}
                                                                                                                                  &\multicolumn{1}{c}{-1,041,674}
                                                                                                                                      & \multicolumn{1}{c}{-1,046,540}\\
\hline
\end{tabular}
}
\begin{footnotesize}
\begin{quote}
{\sl NOTE:  The reported numbers are the posterior means for the parameters, posterior standard deviations are in parentheses  and  effective sample sizes (ESS) are in brackets.
$\log p_{\hat\theta}(y_{1:T})$ is  the  log-likelihood evaluated at the posterior mean of the parameters, and DIC the deviance information criterion. Results are based on 11,000 Gibbs iterations discarding the first 1,000 draws.}
\end{quote}
\end{footnotesize}
\end{table}

\pagebreak
%
\begin{landscape}
\begin{table}\centering
\footnotesize{
\begin{tabular}{lrrcrrcrrcrrcrrcrrcrrcrr}
\multicolumn{24}{c}{\mbox{\small{\sl Table 2. Mean and standard deviation of the predictive Pearson residuals and log-predictive likelihood. }}}\\[0.1cm]\hline\\[0.0cm]
&\multicolumn{11}{c}{Jan 2, 2008 to Dec 31, 2008}&&\multicolumn{11}{c}{ Jun 1, 2015 to May 31, 2016}\\
\cline{2-12}\cline{14-24}\\
\multicolumn{1}{l}{}   & \multicolumn{2}{c}{3F}&& \multicolumn{2}{c}{3F-SV}  && \multicolumn{2}{c}{4F}&& \multicolumn{2}{c}{4F-SV}&&
\multicolumn{2}{c}{3F} && \multicolumn{2}{c}{3F-SV}  && \multicolumn{2}{c}{4F}&& \multicolumn{2}{c}{4F-SV}\\
\cline{2-3}\cline{5-6}\cline{8-9}\cline{11-12}\cline{14-15}\cline{17-18}\cline{20-21}\cline{23-24}\\[-0.2cm]
\multicolumn{1}{l}{Cont.}
&\multicolumn{2}{c}{Mean$\quad$ Sd}&
&\multicolumn{2}{c}{Mean$\quad$ Sd}&
&\multicolumn{2}{c}{Mean$\quad$ Sd}&
&\multicolumn{2}{c}{Mean$\quad$ Sd}&
&\multicolumn{2}{c}{Mean$\quad$ Sd}&
&\multicolumn{2}{c}{Mean$\quad$ Sd}&
&\multicolumn{2}{c}{Mean$\quad$ Sd}&
&\multicolumn{2}{c}{Mean$\quad$ Sd}\\
\hline\\
1   & -0.22  &  1.38   && -0.10      & 1.12     && -0.15    &  1.33   && -0.06    &  1.11  &&-0.29    & 1.21   && -0.29   &1.03    && 0.01   &1.21   &&-0.03  & 1.03 \\
2   & -0.16  &  1.39   && -0.08      & 1.11     && -0.17    &  1.37   && -0.08    &  1.11  &&-0.07    & 1.23   && -0.11   &1.03    &&-0.10   &1.23   &&-0.12  & 1.04 \\
3   & -0.13  &  1.42   && -0.07      & 1.11     && -0.17    &  1.38   && -0.08    &  1.10  && 0.06    & 1.25   &&  0.00   &1.03    &&-0.11   &1.23   &&-0.12  & 1.03 \\
4   & -0.12  &  1.44   &&  -0.06     & 1.10     && -0.16    &  1.39   && -0.08    &  1.10  && 0.14    & 1.25   &&  0.07   &1.03    &&-0.06   &1.23   &&-0.08  & 1.03 \\
5   & -0.11  &  1.46   && -0.05      & 1.10     && -0.14    &  1.40   && -0.06    &  1.10  && 0.16    & 1.25   &&  0.09   &1.03    &&-0.02   &1.23   &&-0.04  & 1.03 \\
6   & -0.11  &  1.48   && -0.04      & 1.10     && -0.13    &  1.42   && -0.04    &  1.10  && 0.13    & 1.25   &&  0.07   &1.03    && 0.00   &1.23   &&-0.02  & 1.03 \\
7   & -0.12  &  1.49   && -0.04      & 1.10     && -0.13    &  1.43   && -0.03    &  1.10  && 0.07    & 1.25   &&  0.03   &1.03    && 0.01   &1.24   &&-0.01  & 1.04 \\
8   & -0.13  &  1.50   &&  -0.04     & 1.10     && -0.13    &  1.44   && -0.03    &  1.10  && 0.01    & 1.24   &&  -0.03  &1.03    && 0.01   &1.24   &&-0.02  & 1.04 \\
9   & -0.15  &  1.51   && -0.05      & 1.10     && -0.13    &  1.45   && -0.03    &  1.11  &&-0.07    & 1.24   && -0.08   &1.03    &&-0.01   &1.23   &&-0.04  & 1.04 \\
10  & -0.16  &  1.51   && -0.05      & 1.10     && -0.13    &  1.45   && -0.03    &  1.11  &&-0.12    & 1.23   && -0.13   &1.03    &&-0.03   &1.22   &&-0.05  & 1.03 \\
11  & -0.17  &  1.51   && -0.06      & 1.10     && -0.14    &  1.45   && -0.04    &  1.11  &&-0.16    & 1.22   && -0.16   &1.04    &&-0.04   &1.22   &&-0.06  & 1.04 \\
12  & -0.17  &  1.52   && -0.06      & 1.10     && -0.14    &  1.46   && -0.04    &  1.10  &&-0.19    & 1.22   && -0.19   &1.04    &&-0.05   &1.21   &&-0.08  & 1.04 \\
13  & -0.18  &  1.52   && -0.06      & 1.10     && -0.14    &  1.45   && -0.04    &  1.10  &&-0.20    & 1.21   && -0.21   &1.05    &&-0.06   &1.21   &&-0.09  & 1.05 \\
14  & -0.18  &  1.51   &&  -0.07     & 1.10     && -0.15    &  1.45   && -0.05    &  1.10  &&-0.21    & 1.20   &&  -0.22  &1.04    &&-0.07   &1.20   &&-0.10  & 1.05 \\
15  & -0.18  &  1.51   && -0.07      & 1.10     && -0.15    &  1.45   && -0.05    &  1.10  &&-0.20    & 1.19   && -0.22   &1.04    &&-0.08   &1.20   &&-0.11  & 1.04 \\
16  & -0.17  &  1.51   && -0.06      & 1.10     && -0.14    &  1.45   && -0.05    &  1.10  &&-0.19    & 1.19   && -0.21   &1.04    &&-0.09   &1.19   &&-0.12  & 1.04 \\
17  & -0.15  &  1.50   && -0.05      & 1.10     && -0.13    &  1.45   && -0.04    &  1.10  &&-0.17    & 1.18   && -0.19   &1.05    &&-0.09   &1.19   &&-0.12  & 1.05 \\
18  & -0.14  &  1.50   && -0.04      & 1.09     && -0.13    &  1.44   && -0.04    &  1.10  &&-0.15    & 1.18   && -0.18   &1.05    &&-0.10   &1.18   &&-0.13  & 1.06 \\
19  & -0.13  &  1.49   && -0.04      & 1.09     && -0.13    &  1.44   && -0.03    &  1.09  &&-0.10    & 1.17   && -0.15   &1.05    &&-0.09   &1.18   &&-0.13  & 1.06 \\
20  & -0.12  &  1.48   &&  -0.03     & 1.09     && -0.12    &  1.44   && -0.02    &  1.09  &&-0.06    & 1.17   &&  -0.10  &1.05    &&-0.08   &1.17   &&-0.12  & 1.05 \\
21  & -0.10  &  1.48   && -0.02      & 1.09     && -0.12    &  1.43   && -0.02    &  1.09  &&-0.00    & 1.16   && -0.05   &1.04    &&-0.06   &1.16   &&-0.10  & 1.05 \\
22  & -0.08  &  1.47   &&  -0.00     & 1.09     && -0.11    &  1.43   && -0.01    &  1.09  && 0.06    & 1.14   &&  0.01   &1.04    &&-0.04   &1.15   &&-0.08  & 1.04 \\
23  & -0.06  &  1.47   && 0.01       & 1.09     && -0.09    &  1.42   &&  0.01    &  1.09  && 0.13    & 1.14   && 0.08    &1.04    &&-0.01   &1.14   &&-0.05  & 1.05 \\
24  & -0.04  & 1.47    && 0.03       & 1.09     && -0.08    &  1.42   &&  0.02    &  1.09  && 0.20    &1.13    && 0.14    &1.04    && 0.02   &1.13   &&-0.02  & 1.05 \\

\multicolumn{24}{c}{}\\[-0.0cm]
&\multicolumn{23}{c}{Log-predictive likelihood}\\
\cline{2-24}\\[-0.2cm]
&\multicolumn{2}{c}{24,853}&&\multicolumn{2}{c}{25,019}&&\multicolumn{2}{c}{26,034}&&\multicolumn{2}{c}{26,195}
         &&\multicolumn{2}{c}{24,420}&&\multicolumn{2}{c}{24,474}&&\multicolumn{2}{c}{27,060}&&\multicolumn{2}{c}{27,130}\\
\hline
\end{tabular}
}
\begin{footnotesize}
\begin{quote}
{\sl }
\end{quote}
\end{footnotesize}
\end{table}
\end{landscape}

\pagebreak
%
\begin{landscape}
\begin{table}\centering
\footnotesize{
\begin{tabular}{lrrcrrcrrcrrcrrcrrcrrcrr}
\multicolumn{24}{c}{\mbox{\small{\sl Table 3. $p$-values of the Ljung-Box test for the predictive Pearson residuals. }}}\\[0.1cm]\hline\\[0.0cm]
&\multicolumn{11}{c}{Jan 2, 2008 to Dec 31, 2008}&&\multicolumn{11}{c}{Jun 1, 2015 to May 31, 2016}\\
\cline{2-12}\cline{14-24}\\
\multicolumn{1}{l}{}   & \multicolumn{2}{c}{3F}&& \multicolumn{2}{c}{3F-SV}  && \multicolumn{2}{c}{4F}&& \multicolumn{2}{c}{4F-SV}&&
\multicolumn{2}{c}{3F} && \multicolumn{2}{c}{3F-SV}  && \multicolumn{2}{c}{4F}&& \multicolumn{2}{c}{4F-SV}\\
\cline{2-3}\cline{5-6}\cline{8-9}\cline{11-12}\cline{14-15}\cline{17-18}\cline{20-21}\cline{23-24}\\[-0.2cm]
\multicolumn{1}{l}{Cont.}
&\multicolumn{1}{c}{$\xi_{it+1}$}&\multicolumn{1}{c}{$\xi_{it+1}^2$}&
&\multicolumn{1}{c}{$\xi_{it+1}$}&\multicolumn{1}{c}{$\xi_{it+1}^2$}&
&\multicolumn{1}{c}{$\xi_{it+1}$}&\multicolumn{1}{c}{$\xi_{it+1}^2$}&
&\multicolumn{1}{c}{$\xi_{it+1}$}&\multicolumn{1}{c}{$\xi_{it+1}^2$}&
&\multicolumn{1}{c}{$\xi_{it+1}$}&\multicolumn{1}{c}{$\xi_{it+1}^2$}&
&\multicolumn{1}{c}{$\xi_{it+1}$}&\multicolumn{1}{c}{$\xi_{it+1}^2$}&
&\multicolumn{1}{c}{$\xi_{it+1}$}&\multicolumn{1}{c}{$\xi_{it+1}^2$}&
&\multicolumn{1}{c}{$\xi_{it+1}$}&\multicolumn{1}{c}{$\xi_{it+1}^2$}\\
\hline\\
1& \cellcolor{gray!70}0.00 & \cellcolor{gray!70} 0.00 &&0.46  &                    0.29  &&\cellcolor{gray!70}0.01& \cellcolor{gray!70}0.00  && 0.48 &                     0.23 && 0.64 &  \cellcolor{gray!70}0.00 &&0.58  & 0.11  &&0.58 &  \cellcolor{gray!70}0.00 && 0.67 &  0.12\\
2& \cellcolor{gray!70}0.01 & \cellcolor{gray!70} 0.00 &&0.48  &                    0.15  &&\cellcolor{gray!70}0.01& \cellcolor{gray!70}0.00  && 0.45 &                     0.17 && 0.58 &  \cellcolor{gray!70}0.00 &&0.52  & 0.35  &&0.52 &  \cellcolor{gray!70}0.00 && 0.58 &  0.22 \\
3& \cellcolor{gray!70}0.01 & \cellcolor{gray!70} 0.00 &&0.50  &                    0.09  &&\cellcolor{gray!70}0.01& \cellcolor{gray!70}0.00  && 0.47 &                     0.11 && 0.54 &  \cellcolor{gray!70}0.00 &&0.47  & 0.37  &&0.41 &  \cellcolor{gray!70}0.00 && 0.50 &  0.19 \\
4& \cellcolor{gray!30}0.01 & \cellcolor{gray!70} 0.00 &&0.54  &                    0.07  &&\cellcolor{gray!70}0.01& \cellcolor{gray!70}0.00  && 0.49 &                     0.07 && 0.50 &  \cellcolor{gray!70}0.00 &&0.41  & 0.39  &&0.29 &  \cellcolor{gray!70}0.00 && 0.38 &  0.18 \\
5& \cellcolor{gray!30}0.02 & \cellcolor{gray!70} 0.00 &&0.56  & \cellcolor{gray!30}0.05  &&\cellcolor{gray!30}0.01& \cellcolor{gray!70}0.00  && 0.53 &  \cellcolor{gray!30}0.04 && 0.40 &  \cellcolor{gray!70}0.00 &&0.30  & 0.38  &&0.21 &  \cellcolor{gray!70}0.00 && 0.28 &  0.15 \\
6& \cellcolor{gray!30}0.02 & \cellcolor{gray!70} 0.00 &&0.58  & \cellcolor{gray!30}0.03  &&\cellcolor{gray!30}0.02& \cellcolor{gray!70}0.00  && 0.55 &  \cellcolor{gray!30}0.03 && 0.34 &  \cellcolor{gray!70}0.00 &&0.24  & 0.38  &&0.22 &  \cellcolor{gray!70}0.00 && 0.25 &  0.17 \\
7& \cellcolor{gray!30}0.03 & \cellcolor{gray!70} 0.00 &&0.60  & \cellcolor{gray!30}0.02  &&\cellcolor{gray!30}0.03& \cellcolor{gray!70}0.00  && 0.59 &  \cellcolor{gray!30}0.02 && 0.30 &  \cellcolor{gray!70}0.00 &&0.22  & 0.42  &&0.22 &  \cellcolor{gray!70}0.00 && 0.26 &  0.21 \\
8& \cellcolor{gray!30}0.04 & \cellcolor{gray!70} 0.00 &&0.63  & \cellcolor{gray!30}0.02  &&\cellcolor{gray!30}0.04& \cellcolor{gray!70}0.00  && 0.62 &  \cellcolor{gray!30}0.02 && 0.30 &  \cellcolor{gray!70}0.00 &&0.21  & 0.38  &&0.22 &  \cellcolor{gray!70}0.00 && 0.26 &  0.26 \\
9& \cellcolor{gray!30}0.05 & \cellcolor{gray!70} 0.00 &&0.64  & \cellcolor{gray!30}0.02  &&\cellcolor{gray!30}0.04& \cellcolor{gray!70}0.00  && 0.65 &  \cellcolor{gray!30}0.02 && 0.26 &  \cellcolor{gray!70}0.00 &&0.20  & 0.39  &&0.18 &  \cellcolor{gray!70}0.00 && 0.23 &  0.27 \\
10&                   0.05 & \cellcolor{gray!70} 0.00 &&0.65  & \cellcolor{gray!30}0.01  &&                   0.06& \cellcolor{gray!70}0.00  && 0.66 &  \cellcolor{gray!30}0.01 && 0.21 &  \cellcolor{gray!70}0.00 &&0.20  & 0.34  &&0.16 &  \cellcolor{gray!70}0.00 && 0.22 &  0.26 \\
11&                   0.06 & \cellcolor{gray!70} 0.00 &&0.67  & \cellcolor{gray!30}0.01  &&                   0.07& \cellcolor{gray!70}0.00  && 0.68 &  \cellcolor{gray!30}0.01 && 0.20 &  \cellcolor{gray!70}0.00 &&0.20  & 0.40  &&0.16 &  \cellcolor{gray!70}0.00 && 0.22 &  0.26 \\
12&                   0.07 & \cellcolor{gray!70} 0.00 &&0.68  & \cellcolor{gray!30}0.01  &&                   0.08& \cellcolor{gray!70}0.00  && 0.71 &  \cellcolor{gray!30}0.01 && 0.23 &  \cellcolor{gray!70}0.00 &&0.22  & 0.37  &&0.19 &  \cellcolor{gray!70}0.00 && 0.26 &  0.33 \\
13&                   0.08 & \cellcolor{gray!70} 0.00 &&0.70  & \cellcolor{gray!70}0.01  &&                   0.08& \cellcolor{gray!70}0.00  && 0.74 &  \cellcolor{gray!30}0.01 && 0.27 &  \cellcolor{gray!70}0.00 &&0.25  & 0.37  &&0.22 &  \cellcolor{gray!70}0.00 && 0.29 &  0.32 \\
14&                   0.09 & \cellcolor{gray!70} 0.00 &&0.73  & \cellcolor{gray!70}0.01  &&                   0.10& \cellcolor{gray!70}0.00  && 0.75 &  \cellcolor{gray!70}0.01 && 0.25 &  \cellcolor{gray!70}0.00 &&0.27  & 0.37  &&0.21 &  \cellcolor{gray!70}0.00 && 0.31 &  0.32 \\
15&                   0.11 & \cellcolor{gray!70} 0.00 &&0.75  & \cellcolor{gray!30}0.01  &&                   0.11& \cellcolor{gray!70}0.00  && 0.76 &  \cellcolor{gray!70}0.01 && 0.22 &  \cellcolor{gray!70}0.00 &&0.23  & 0.33  &&0.18 &  \cellcolor{gray!70}0.00 && 0.26 &  0.29 \\
16&                   0.11 & \cellcolor{gray!70} 0.00 &&0.75  & \cellcolor{gray!70}0.01  &&                   0.13& \cellcolor{gray!70}0.00  && 0.78 &  \cellcolor{gray!70}0.01 && 0.21 &  \cellcolor{gray!70}0.00 &&0.23  & 0.23  &&0.16 &  \cellcolor{gray!70}0.00 && 0.24 &  0.25 \\
17&                   0.13 & \cellcolor{gray!70} 0.00 &&0.76  & \cellcolor{gray!70}0.01  &&                   0.15& \cellcolor{gray!70}0.00  && 0.79 &  \cellcolor{gray!70}0.01 && 0.26 &  \cellcolor{gray!70}0.00 &&0.25  & 0.29  &&0.19 &  \cellcolor{gray!70}0.00 && 0.27 &  0.26 \\
18&                   0.15 & \cellcolor{gray!70} 0.00 &&0.77  & \cellcolor{gray!70}0.01  &&                   0.15& \cellcolor{gray!70}0.00  && 0.80 &  \cellcolor{gray!70}0.01 && 0.33 &  \cellcolor{gray!70}0.00 &&0.28  & 0.38  &&0.23 &  \cellcolor{gray!70}0.00 && 0.32 &  0.33 \\
19&                   0.16 & \cellcolor{gray!70} 0.00 &&0.79  & \cellcolor{gray!70}0.01  &&                   0.17& \cellcolor{gray!70}0.00  && 0.80 &  \cellcolor{gray!70}0.01 && 0.36 &  \cellcolor{gray!70}0.00 &&0.28  & 0.52  &&0.26 &  \cellcolor{gray!70}0.00 && 0.33 &  0.39 \\
20&                   0.18 & \cellcolor{gray!70} 0.00 &&0.79  & \cellcolor{gray!70}0.01  &&                   0.19& \cellcolor{gray!70}0.00  && 0.81 &  \cellcolor{gray!70}0.01 && 0.38 &  \cellcolor{gray!70}0.00 &&0.29  & 0.49  &&0.24 &  \cellcolor{gray!70}0.00 && 0.34 &  0.37 \\
21&                   0.20 & \cellcolor{gray!70} 0.00 &&0.81  & \cellcolor{gray!70}0.01  &&                   0.21& \cellcolor{gray!70}0.00  && 0.82 &  \cellcolor{gray!70}0.01 && 0.33 &  \cellcolor{gray!70}0.00 &&0.24  & 0.54  &&0.19 &  \cellcolor{gray!70}0.00 && 0.29 &  0.28 \\
22&                   0.23 & \cellcolor{gray!70} 0.00 &&0.80  & \cellcolor{gray!70}0.01  &&                   0.22& \cellcolor{gray!70}0.00  && 0.81 &  \cellcolor{gray!70}0.01 && 0.24 &  \cellcolor{gray!70}0.00 &&0.19  & 0.56  &&0.14 &  \cellcolor{gray!70}0.00 && 0.23 &  0.29 \\
23&                   0.26 & \cellcolor{gray!70} 0.00 &&0.81  & \cellcolor{gray!70}0.01  &&                   0.25& \cellcolor{gray!70}0.00  && 0.82 &  \cellcolor{gray!70}0.01 && 0.24 &  \cellcolor{gray!70}0.00 &&0.16  & 0.70  &&0.14 &  \cellcolor{gray!70}0.00 && 0.20 &  0.45 \\
24&                   0.28 & \cellcolor{gray!70} 0.00 &&0.80  & \cellcolor{gray!70}0.01  &&                   0.29& \cellcolor{gray!70}0.00  && 0.81 &  \cellcolor{gray!30}0.01 && 0.27 &  \cellcolor{gray!70}0.00 &&0.18  & 0.75  &&0.17 &  \cellcolor{gray!70}0.00 && 0.27 &  0.49 \\
\multicolumn{24}{c}{}\\[-0.3cm]
\hline
\end{tabular}
}
\begin{footnotesize}
\begin{quote}
{\sl NOTE: The Ljung-Box test statistic for the predictive Pearson residuals $\xi_{it+1}$ and $\xi_{it+1}^2$   is computed including 10 lags.
Light grey-shaded cells indicate significance at the 5\% level and dark-shaded cells significance at the 1\% level.}
\end{quote}
\end{footnotesize}
\end{table}
\end{landscape}

\pagebreak
%
%

\begin{table}[h!]\centering
\footnotesize{
\begin{tabular}{l.....}
\multicolumn{6}{c}{\mbox{\small{\sl Table 4. Root-mean-squared forecast errors.  }}}\\[0.1cm]\hline\\[-0.3cm]

\multicolumn{6}{l}{ Jan 2, 2008 to Dec 31, 2008}\\[0.0cm]

\multicolumn{1}{l}{} & \multicolumn{1}{c}{}    & \multicolumn{1}{c}{}      &\multicolumn{1}{c}{ }  &\multicolumn{1}{c}{}&\multicolumn{1}{c}{Random}\\
\multicolumn{1}{l}{} & \multicolumn{1}{c}{3F}  & \multicolumn{1}{c}{3F-SV} &\multicolumn{1}{c}{4F} &\multicolumn{1}{c}{4F-SV}&\multicolumn{1}{c}{walk}\\
\hline\\
Contract 1-8               &  0.0320   &   0.0320   &   0.0319   & 0.0319  & 0.0316   \\
Contract 9-16              &  0.0285   &   0.0285   &   0.0284   & 0.0284  & 0.0281   \\
Contract 17-24             &  0.0261   &   0.0261   &   0.0261   & 0.0261  & 0.0259   \\
All Contracts     &  0.0290   &   0.0290   &   0.0289   & 0.0289  & 0.0286\\[0.0cm]\hline\\[-0.3cm] 

\multicolumn{6}{l}{Jun 1, 2015 to May 31, 2016}\\[0.0cm]
\multicolumn{1}{l}{} & \multicolumn{1}{c}{}    & \multicolumn{1}{c}{}      &\multicolumn{1}{c}{ }  &\multicolumn{1}{c}{}&\multicolumn{1}{c}{Random}\\
\multicolumn{1}{l}{} & \multicolumn{1}{c}{3F}  & \multicolumn{1}{c}{3F-SV} &\multicolumn{1}{c}{4F} &\multicolumn{1}{c}{4F-SV}&\multicolumn{1}{c}{walk}\\
\hline\\
Contract 1-8               &  0.0269   &   0.0271   &   0.0269   & 0.0268  & 0.0264   \\
Contract 9-16              &  0.0224   &   0.0226   &   0.0225   & 0.0224  & 0.0221   \\
Contract 17-24             &  0.0198   &   0.0199   &   0.0200   & 0.0197  & 0.0196   \\
All Contracts     &  0.0232   &   0.0234   &   0.0233   & 0.0232  & 0.0229\\
\hline
\end{tabular}
}


%
%
\vspace{2cm}
\centering
\footnotesize{
\begin{tabular}{lcccccccccccccc}
\multicolumn{14}{c}{\mbox{\small{\sl Table 5. Value at Risk forecasting results for an equally weighted portfolio}}}\\[0.1cm]\hline\\[-0.2cm]
\multicolumn{1}{l}{}   & \multicolumn{1}{c}{}&\multicolumn{1}{c}{}  & \multicolumn{1}{c}{}&\multicolumn{1}{c}{}&\multicolumn{1}{c}{VAR}&\multicolumn{1}{c}{VAR}&&
                         \multicolumn{1}{c}{}&\multicolumn{1}{c}{}  & \multicolumn{1}{c}{}&\multicolumn{1}{c}{}&\multicolumn{1}{c}{VAR}&\multicolumn{1}{c}{VAR}\\
\multicolumn{1}{l}{}   & \multicolumn{1}{c}{3F}  &\multicolumn{1}{c}{3F-SV} & \multicolumn{1}{c}{4F}&\multicolumn{1}{c}{4F-SV}&\multicolumn{1}{c}{-3F}&\multicolumn{1}{c}{-4F}&&
                         \multicolumn{1}{c}{3F}  &\multicolumn{1}{c}{3F-SV}  & \multicolumn{1}{c}{4F}&\multicolumn{1}{c}{4F-SV}&\multicolumn{1}{c}{-3F}&\multicolumn{1}{c}{-4F}\\ \hline\\[0.0cm]
&\multicolumn{6}{c}{Jan 2, 2008 to Dec 31, 2008}&&\multicolumn{6}{c}{Jun 1, 2015 to May 31, 2016}\\
\cline{2-7}\cline{9-14}\\[-0.2cm]

\multicolumn{14}{l}{Hit rate}\\[0.2cm]
1\%               & 0.08 & 0.02 & 0.08& 0.02&0.07&0.10 && 0.02 & 0.02& 0.02& 0.02 &0.02& 0.02 \\
5\%               & 0.15 & 0.06 & 0.14& 0.06&0.14&0.18 && 0.07 & 0.04& 0.07& 0.04 &0.05& 0.08 \\
10\%\qquad        & 0.20 & 0.13 & 0.19& 0.14&0.19&0.23 && 0.15 & 0.11& 0.14& 0.10 &0.11& 0.16 \\

\multicolumn{14}{c}{}\\[0.0cm]
\multicolumn{14}{l}{Ind.}\\[0.2cm]
1\%                & 0.34 & 0.66 & 0.21& 0.66&0.44&0.40 && 0.73 & 0.73& 0.66& 0.60 &0.73& 0.66 \\
5\%                & 0.82 & 0.80 & 0.93& 0.98&0.93&0.52 && 0.43 & 0.32& 0.43& 0.20 &0.28& 0.73 \\
10\%               & 0.68 & \cellcolor{gray!30} 0.04 & 0.57& 0.30&0.57&0.76 && 0.38 & 0.46& 0.55& 0.73 &0.58& 0.19 \\
\multicolumn{14}{c}{}\\[0.0cm]
\multicolumn{14}{l}{UC}\\[0.2cm]
1\%   &  \cellcolor{gray!70} 0.00 & 0.16 &  \cellcolor{gray!70} 0.00& 0.16& \cellcolor{gray!70} 0.00&  \cellcolor{gray!70} 0.00 &
                      &  \cellcolor{gray!00} 0.39 & 0.17& \cellcolor{gray!00} 0.17& 0.17 &0.39& 0.17 \\
5\%   &  \cellcolor{gray!70} 0.00 & 0.68 &  \cellcolor{gray!70} 0.00& 0.34& \cellcolor{gray!70} 0.00&  \cellcolor{gray!70} 0.00 &
                      &       0.23 & 0.91& 0.23 & 0.64 &0.86&\cellcolor{gray!30} 0.05 \\
10\%  &  \cellcolor{gray!70} 0.00 & 0.16 &  \cellcolor{gray!70} 0.00&\cellcolor{gray!30} 0.05& \cellcolor{gray!70} 0.00&  \cellcolor{gray!70} 0.00 &
                      &  \cellcolor{gray!30} 0.01 & 0.56& \cellcolor{gray!30} 0.03& 0.97 &0.56& \cellcolor{gray!70} 0.00\\
\multicolumn{14}{c}{}\\[0.0cm]
\multicolumn{14}{l}{CC}\\[0.2cm]
1\%   &  \cellcolor{gray!70} 0.00 & 0.35 &  \cellcolor{gray!70} 0.00& 0.35& \cellcolor{gray!70} 0.00& \cellcolor{gray!70} 0.00 &
                      & 0.65 & 0.65& 0.35& 0.35 &0.65 &      0.35 \\
5\%   & \cellcolor{gray!70} 0.00 & 0.89 & \cellcolor{gray!70} 0.00& 0.63& \cellcolor{gray!70} 0.00& \cellcolor{gray!70}  0.00 &
                      & 0.35 & 0.55& 0.35& 0.55 &0.55&      0.13 \\
10\%  & \cellcolor{gray!70}  0.00 &\cellcolor{gray!30} 0.05 &  \cellcolor{gray!70} 0.00& 0.08& \cellcolor{gray!70} 0.00& \cellcolor{gray!70} 0.00 &
                      &\cellcolor{gray!30} 0.03 & 0.64& 0.08& 0.94 &0.73&  \cellcolor{gray!70} 0.00 \\
\multicolumn{14}{c}{}\\[-0.3cm]
\hline
\end{tabular}
}
\begin{footnotesize}
\begin{quote}
{\sl NOTE: Reported numbers are the hit rate for the predicted 1\%, 5\% and 10\% VaR,  and $p$-values for the test of independence (Ind.), unconditional coverage (UC) and conditional coverage (CC).
Light grey-shaded cells indicate significance at the 5\% level and dark-shaded cells significance at the 1\% level.}
\end{quote}
\end{footnotesize}
\end{table}

\pagebreak

%
%
\begin{table}\centering
\footnotesize{
\begin{tabular}{lcccccccccccccc}
\multicolumn{14}{c}{\mbox{\small{\sl Table 6. Value at Risk forecasting results for a bull spread portfolio}}}\\[0.1cm]\hline\\[-0.2cm]
\multicolumn{1}{l}{}   & \multicolumn{1}{c}{}&\multicolumn{1}{c}{}  & \multicolumn{1}{c}{}&\multicolumn{1}{c}{}&\multicolumn{1}{c}{VAR}&\multicolumn{1}{c}{VAR}&&
                         \multicolumn{1}{c}{}&\multicolumn{1}{c}{}   & \multicolumn{1}{c}{}&\multicolumn{1}{c}{}&\multicolumn{1}{c}{VAR}&\multicolumn{1}{c}{VAR}\\
\multicolumn{1}{l}{}   & \multicolumn{1}{c}{3F}  &\multicolumn{1}{c}{3F-SV}  & \multicolumn{1}{c}{4F}&\multicolumn{1}{c}{4F-SV}&\multicolumn{1}{c}{-3F}&\multicolumn{1}{c}{-4F}&&
                         \multicolumn{1}{c}{3F}  &\multicolumn{1}{c}{3F-SV}  & \multicolumn{1}{c}{4F}&\multicolumn{1}{c}{4F-SV}&\multicolumn{1}{c}{-3F}&\multicolumn{1}{c}{-4F}\\ \hline\\[0.0cm]
&\multicolumn{6}{c}{Jan 2, 2008 to Dec 31, 2008}&&\multicolumn{6}{c}{Jun 1, 2015 to May 31, 2016}\\
\cline{2-7}\cline{9-14}\\[-0.2cm]
\multicolumn{14}{l}{Hit rate}\\[0.2cm]
1\%                 & 0.04 & 0.04 & 0.01   & 0.00&0.06&0.02 && 0.06 & 0.06& 0.00& 0.01 &0.11& 0.02 \\
5\%                 & 0.06 & 0.08 & 0.03& 0.05&0.09&0.04 && 0.14 & 0.19& 0.02& 0.03 &0.25& 0.05 \\
10\% \qquad         & 0.09 & 0.10 & 0.05& 0.08&0.10&0.07 && 0.23 & 0.27& 0.06& 0.05 &0.36& 0.09 \\
\multicolumn{14}{c}{}\\[0.0cm]
\multicolumn{14}{l}{Ind.}\\[0.2cm]
1\%   & \cellcolor{gray!70} 0.00 & \cellcolor{gray!70} 0.00 &\cellcolor{gray!30} 0.01& \cellcolor{gray!00} 0.94&\cellcolor{gray!70} 0.00&                    0.06 &
                        &                     0.05 &                     0.91 & 0.94&                     0.87&                    0.06&                    0.73 \\
5\%   & \cellcolor{gray!70} 0.00 & \cellcolor{gray!70} 0.00 & 0.12&                     0.49&\cellcolor{gray!70} 0.00&                    0.40 &
                        & \cellcolor{gray!70} 0.00 & \cellcolor{gray!70} 0.00 & 0.66&                     0.48&\cellcolor{gray!70} 0.01&                    0.24 \\
10\%  & \cellcolor{gray!70} 0.00 & \cellcolor{gray!70} 0.00 & 0.59& \cellcolor{gray!70} 0.00&\cellcolor{gray!70} 0.00&\cellcolor{gray!70} 0.00 &
                        & \cellcolor{gray!70 }0.00 & \cellcolor{gray!70} 0.00 & 0.20&                     0.16&\cellcolor{gray!30} 0.03&                    0.36 \\
\multicolumn{14}{c}{}\\[0.0cm]
\multicolumn{14}{l}{UC}\\[0.2cm]
1\%   &  \cellcolor{gray!70}  0.00 & \cellcolor{gray!70} 0.00 &                      0.76& 0.28 & \cellcolor{gray!70} 0.00&                     0.16 &
                      & \cellcolor{gray!70}   0.00 & \cellcolor{gray!70} 0.00 &                      0.27& 0.73 & \cellcolor{gray!70} 0.00&                     0.39 \\
5\%   &                       0.49 &                     0.08 &  \cellcolor{gray!30} 0.04& 0.65 & \cellcolor{gray!30} 0.01&                     0.45 &
                      &   \cellcolor{gray!70} 0.00 & \cellcolor{gray!70} 0.00 &                      0.13& 0.16 & \cellcolor{gray!70} 0.00&                     0.91 \\
10\%  &                       0.38 &                     0.82 &  \cellcolor{gray!70} 0.00& 0.12 &                     0.82& \cellcolor{gray!30} 0.04 &
                      &   \cellcolor{gray!70} 0.00 & \cellcolor{gray!70} 0.00 &  \cellcolor{gray!30} 0.01& \cellcolor{gray!70}0.01 &\cellcolor{gray!70} 0.00 & 0.64 \\
\multicolumn{14}{c}{}\\[0.0cm]
\multicolumn{14}{l}{CC}\\[0.2cm]
1\%   &  \cellcolor{gray!70} 0.00 & \cellcolor{gray!70} 0.00 & \cellcolor{gray!30} 0.04& 0.55& \cellcolor{gray!70} 0.00& 0.06 &
                      &  \cellcolor{gray!70} 0.00 & \cellcolor{gray!70} 0.00 & 0.55& 0.93 & \cellcolor{gray!70}0.00&      0.65 \\
5\%   &  \cellcolor{gray!70} 0.00 & \cellcolor{gray!70} 0.00 &  \cellcolor{gray!30} 0.03& 0.71& \cellcolor{gray!70} 0.00& 0.52 &
                      &  \cellcolor{gray!70} 0.00 & \cellcolor{gray!70} 0.00 &  \cellcolor{gray!30} 0.04& 0.28 &\cellcolor{gray!70} 0.00&      0.50 \\
10\%  &  \cellcolor{gray!70} 0.00 & \cellcolor{gray!70} 0.00 & \cellcolor{gray!70} 0.01& \cellcolor{gray!70} 0.00& \cellcolor{gray!70} 0.00& \cellcolor{gray!70} 0.00 &
                      & \cellcolor{gray!70}  0.00 & \cellcolor{gray!70} 0.00 & \cellcolor{gray!30} 0.02& \cellcolor{gray!70} 0.01& \cellcolor{gray!70} 0.00&      0.59 \\
\multicolumn{14}{c}{}\\[-0.3cm]
\hline
\end{tabular}
}
\begin{footnotesize}
\begin{quote}
{\sl NOTE: Reported numbers are the hit rate for the predicted 1\%, 5\% and 10\% VaR, and $p$-values for the test of independence (Ind.), unconditional coverage (UC) and conditional coverage (CC). Light grey-shaded cells
indicate significance at the 5\% level and dark-shaded cells significance at the 1\% level.}
\end{quote}
\end{footnotesize}
\end{table}

\pagebreak


%
%

\pagebreak
%
%

\pagebreak
%
%
\begin{landscape}
\begin{figure}[!htb]
\centering
\includegraphics[width=1.0\textwidth,angle=-90]{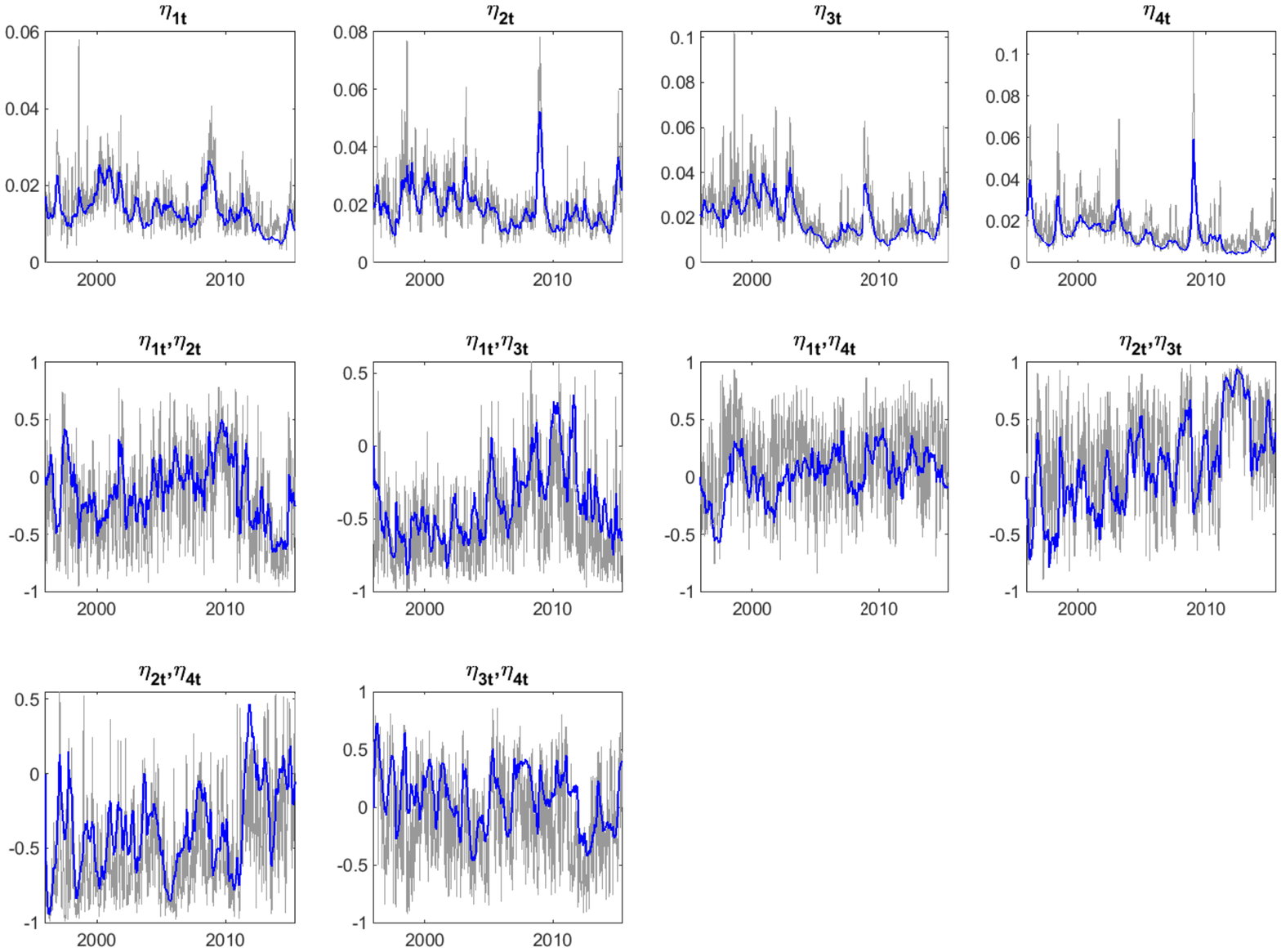}\\[-1.4cm]
\captionsetup{font=small}
\caption{Blue lines: Smoothed estimates for the standard deviations of the factor innovations  $(\eta_{1t},\ldots,\eta_{4t})$ (first row) and their  correlations (second and third row) obtained under the 4F-SV model. Grey lines: two-sided rolling-window estimates of the standard deviations  and correlations obtained from the factors extracted by cross-section least squares.\label{fig:vola_betas}}
\end{figure}
\end{landscape}

\pagebreak
%
%
\begin{figure}[!htb]
\hspace*{-2.0cm}
\includegraphics[width=0.85\textwidth,angle=-90]{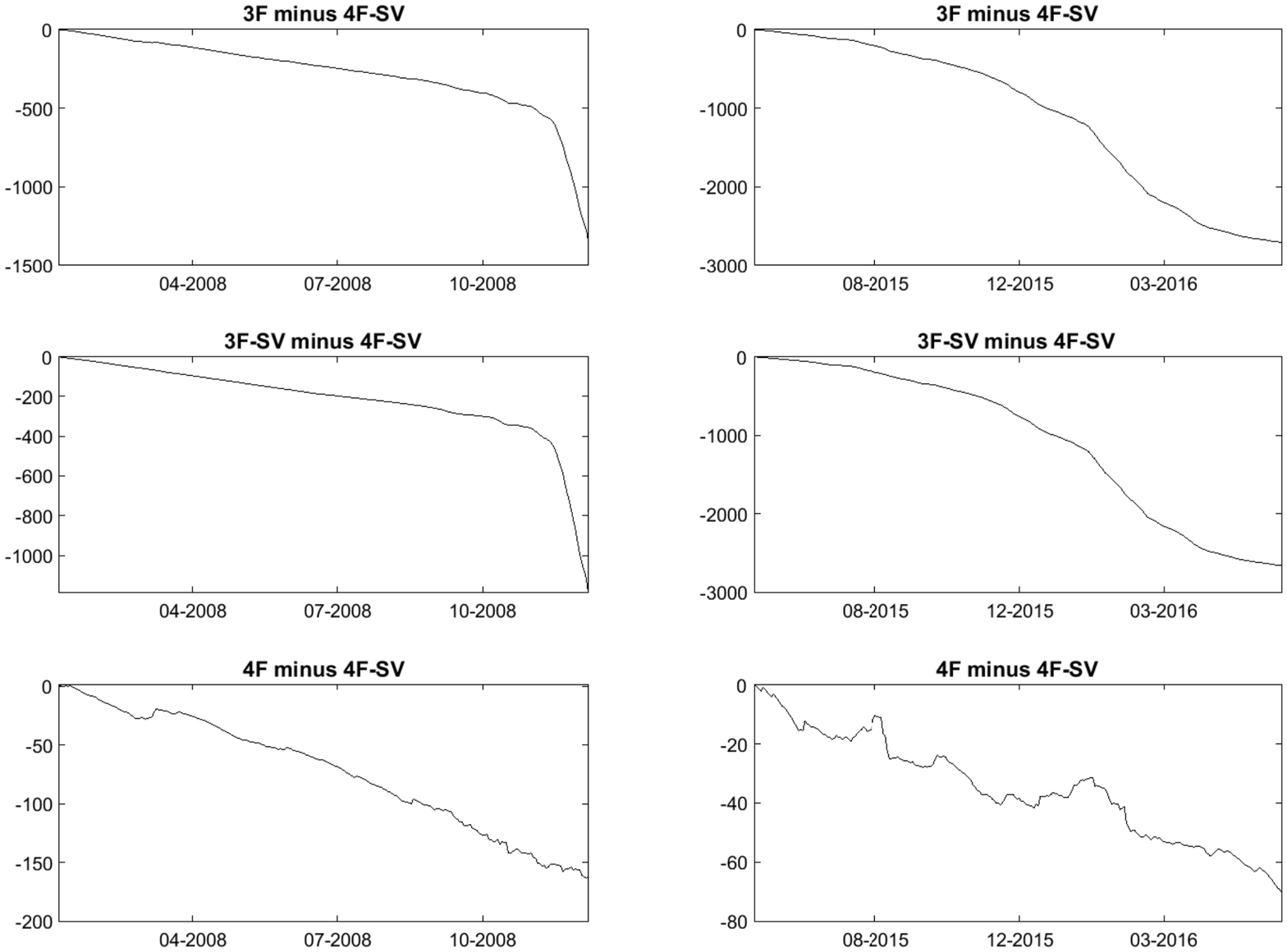}\\[-1.4cm]
\captionsetup{font=small}
\caption{Accumulated log-predictive densities in terms of the difference to the accumulated log-predictive densities of the 4F-SV model  for the out-of-sample period from Jan 2, 2008 to Dec 31, 2008 (left panels) and  the out-of-sample period from Jun 1, 2015 to May 31, 2016 (right panels).\label{fig:pred_dens}}
\end{figure}

\appendix

\subsection*{A1: Derivation of the integrated likelihood in Equation (16)}
In order to obtain the integrated likelihood $f_\theta(\beta_{1:T})= \int f_\theta(\beta_{1:T}|H_{1:T})f_\theta(H_{1:T})dH_{1:T}$  under the multivariate Wishart SV model for $\beta_t$ as defined by Equations (5)-(7)
it is sequentially factorized as
\begin{align}\label{eq-app:integr-eta-likelihood}
f_\theta(\beta_{1:T})=\prod_{t=1}^T f_\theta(\beta_t|\beta_{0:t-1}),\quad\mbox{with}\quad f_\theta(\beta_t|\beta_{0:t-1})=\int f_\theta(\beta_t|H_t,\beta_{0:t-1})f_\theta(H_t|\beta_{0:t-1})dH_t,
\end{align}
where $ f_\theta(\beta_t|H_t,\beta_{0:t-1})= f_\theta(\beta_t|H_t,\beta_{t-1})$ is the conditional ${\cal N}(\alpha+\Phi\beta_{t-1},H_t^{-1})$ density for $\beta_t$ given  $(\beta_{t-1}, H_t)$, and $f_\theta(H_t|\beta_{0:t-1})$ is the one-step ahead predictive density for the state $H_t$.
As shown in Windle and Carvalho (2014, Proof of  Proposition 1), this predictive density for $H_t$ is a  ${\cal W}_m(\nu,(\gamma \Sigma_{t-1})^{-1})$-density with $\Sigma_t=\eta_t\eta_t'+\gamma\Sigma_{t-1}$ and $\eta_t=(\beta_t-\alpha-\Phi\beta_{t-1})$. Thus, $f_\theta(\beta_t|\beta_{0:t-1})$ in Equation (\ref{eq-app:integr-eta-likelihood}) obtains by combining the Gaussian density for $\beta_t$ given $(\beta_{t-1}, H_t)$ with  a Wishart density for $H_t$, which  defines the associated conjugate distribution. Hence, its closed-form formula results from  standard Gaussian-Wishart algebra, from which it follows that
\begin{align}
f_\theta(\beta_t|\beta_{0:t-1})&=\int \underbrace{\frac{1}{(2\pi)^{m/2}}|H_t|^{1/2}\exp\Big\{-\frac{1}{2} \mbox{tr}(\eta_t\eta_t'H_t) \Big\}}_{{\cal N}(0 ,H_t^{-1})} \\
&\qquad \times \underbrace{\frac{1}{2^{m\nu/2}\Gamma_m(\nu/2)}| \gamma \Sigma_{t-1}|^{\nu/2} | H_t|^{(\nu-m-1)/2}\exp\Big\{-\frac{1}{2}\mbox{tr}(\gamma\Sigma_{t-1}H_t) \Big\}}_{{\cal W}_m(\nu,(\gamma \Sigma_{t-1})^{-1})}
    dH_t \nonumber\\[0.2cm]
&= \frac{\Gamma_m((\nu+1)/2)}{\pi^{m/2}\Gamma_m(\nu/2)}|\gamma\Sigma_{t-1}|^{\nu/2} |\eta_t\eta_t'+\gamma\Sigma_{t-1}|^{-(\nu+1)/2}    \\[0.2cm]
&= \frac{\Gamma((\nu+1)/2)}{\pi^{m/2}\Gamma((\nu-m+1)/2)} |\gamma\Sigma_{t-1}|^{-1/2}[1+\eta_t'(\gamma\Sigma_{t-1})^{-1}\eta_t ]^{-(\nu+1)/2},\label{eq-app:t-density}
\end{align}
where $\Gamma_m(\cdot)$ denotes the multivariate  Gamma function. The function in Equation (\ref{eq-app:t-density}) is a multivariate $t$-density  for $\beta_t$ given $\beta_{0:t-1}$ with $\nu-m+1$ degrees of freedom, location parameter $\alpha+\Phi\beta_{t-1}$ and scale matrix $\gamma\Sigma_{t-1}$    (see, e.g., Gupta and Nagar, 2000, p. 334). Accounting for the restriction on $\gamma$ and $\nu$  in Equation (9), 
yields $f_\theta(\beta_t|\beta_{0:t-1})$  as given by Equation (16). (This result for the integrated likelihood $f_\theta(\beta_{1:T})$ is closely related to that of Windle and Carvalho (2014) in their Proposition 3 which provides the analytical form of the integrated likelihood  of the outer products of the Gaussian factor innovations $f_\theta(\{\eta_t\eta_t'\}_{t=1}^T)$.)

\subsection*{A2: SMC approximation of the data likelihood in Equation (18)}
Under the factor SSM model the data likelihood (as reproduced from Equation 18) is given by
\begin{align}\label{eq-app:likelihood-data-1}
f_\theta(y_{1:T})=\int f_{\theta}(y_{1:T}|\beta_{1:T}) f_{\theta}(\beta_{1:T}|H_{1:T})f_\theta(H_{1:T})dH_{1:T}d\beta_{1:T}.
\end{align}
For its evaluation we use a Rao-Blackwellised SMC algorithm (Doucet and Johansen, 2009, Sec.~4.6), which exploits that according to Equations (13) and (16) the precision matrices $H_{1:T}$  can be integrated out  analytically. This
simplifies  the  likelihood integral (\ref{eq-app:likelihood-data-1})  to the lower dimensional integral
\begin{align}\label{eq-app:likelihood-data-2}
f_\theta(y_{1:T})=\int f_{\theta}(y_{1:T}|\beta_{1:T}) f_{\theta}(\beta_{1:T}) d\beta_{1:T}=\int \prod_{t=1}^T f_\theta(y_t|\beta_t)f_\theta(\beta_t|\beta_{0:t-1}) d\beta_{1:T},
\end{align}
where  $f_\theta(y_t|\beta_t)=f_{{\cal N}}(y_t|Z_t'\beta_t,\sigma_y^2I_N)$ is the Gaussian
density as defined by Equation (3), and $f_\theta(\beta_t|\beta_{0:t-1})=f_{{\cal T}}(\beta_t|\alpha+ \beta_{t-1}, \gamma\Sigma_t,\nu-m+1)$ is the $t$-density given in Equation (16)  with $\gamma=(\nu-m-1)/(\nu-m)$.
Our proposed Rao-Blackwellised  SMC algorithm  then consists of applying a standard SMC to the simplified likelihood integral (\ref{eq-app:likelihood-data-2}) which represents the likelihood of a SSM with a Gaussian measurement density and student-$t$ state transition density. According to the Rao-Blackwell Theorem (Robert and Casella, 2004), such  an SMC eliminating a subset of states in the likelihood integral
by analytical integration, can be expected to increase the precision of the  SMC likelihood estimate relative to a `brute-force' SMC application to the initial likelihood integral.

An SMC algorithm for estimating the likelihood (\ref{eq-app:likelihood-data-2}) produces MC-estimates for the sequence of the period-$t$ likelihood contributions $f_\theta(y_t|y_{1:t-1})$ by sequentially importance sampling (IS) and re-sampling using  a sequence of IS densities $q_t(\beta_t)$ for the states $\beta_t$. For a given value of $\theta$ the standard SMC algorithm proceeds as follows (for a detailed treatment of SMC procedures, see Doucet and Johansen, 2009, Capp\'{e} et al., 2007):

{\sl For period $t=1$:} Sample  $\beta_1^{l}\sim q_1(\beta_1)$ for $l=1,\ldots,L$  and compute the period-1 likelihood estimate as
\begin{align}
\hat f_\theta(y_1)= \frac{1}{L}\sum_{l=1}^L w_1^l,\quad\mbox{where}\quad  w_1^l=\frac{f_\theta(y_1|\beta_1^l) f_\theta(\beta_1^l|\beta_0) }{q_1(\beta_1^l)}.
\end{align}
Then compute the normalized IS weights $W_1^{l}=w_{1}^l/(\sum_{\ell=1}^L w_1^{\ell} )$ and re-sample  by drawing  $\bar \beta_1^l$ for $l=1,\ldots,L$  from the set $\{\beta_1^{l} \}_{l=1}^L$ according to their normalized IS weights $\{W_1^{l}\}_{l=1}^L$.

{\sl For periods $t=2,\ldots, T$:} Sample  $\beta_t^{l}\sim q_t(\beta_t)$ for  $l=1,\ldots,L$, set $\beta_{1:t}^l=(\beta_t^l,\bar \beta_{0:t-1}^l)$, and compute the period-t likelihood estimate as
\begin{align}
\hat f_\theta(y_t|y_{1:t-1})= \frac{1}{L}\sum_{l=1}^L w_t^l,\quad\mbox{where}\quad w_t^l=\frac{f_\theta(y_t|\beta_t^l) f_\theta(\beta_t^l|\beta_{0:t-1}^l)}{q_t(\beta_t^l)}.
\end{align}
Then normalize the IS weights  $W_t^{l}=w_{t}^l/(\sum_{\ell=1}^L w_t^{\ell} )$ and re-sample  by drawing  $\bar \beta_{1:t}^l$ for  $l=1,\ldots,L$  from the set $\{\beta_{1:t}^{l} \}_{l=1}^L$ according to their normalized IS weights $\{W_t^{l}\}_{l=1}^L$.

The SMC estimate for the full data likelihood (\ref{eq-app:likelihood-data-2}) which  results from
this algorithm is computed as  $\hat f_\theta(y_{1:T})=[\prod_{t=2}^T\hat f_\theta(y_t|y_{1:t-1})]\hat f_{\theta}(y_1)$.

A standard selection for the SMC-IS densities are the state-transition densities, i.e. $q_t(\beta_t)=f_\theta(\beta_t|\beta_{0:t-1})$ (Gordon et al., 1993). However, under our fitted factor SSM model the measurement density $f_\theta(y_t|\beta_t)=f_{{\cal N}}(y_t|Z_t'\beta_t,\sigma_y^2I_N)$ is highly informative about the states $\beta_t$ so that it is as function in $Z_t'\beta_t$ strongly peaked  at the observed measurements $y_t$.  This is a scenario where an SMC using the state transitions as IS densities
typically  suffers from a large variance in the resulting  IS weights,
which leads to  fairly poor SMC likelihood estimates (Capp\'{e} et al., 2007).
In order to address this problem  we use IS densities which are constructed  such that they closely match period-by-period the location and the shape of the measurement densities. Specifically, we use
the following student-$t$ IS densities
\begin{align}\label{eq-app:IS-density}
q_t(\beta_t)= f_{{\cal T}}(\beta_t|\beta_{t}^*,\Sigma_t^*,\nu^*),
\end{align}
where the location vector $\beta_t^*$ is taken to be an MCMC estimate for the  posterior mean of $\beta_t$ given $\theta$, $\mbox{E}(\beta_t|y_{1:T},\theta)$, and the scale matrix $\Sigma_t^*$  an MCMC estimate for the corresponding posterior covariance matrix $\mbox{Var}(\beta_t|y_{1:T},\theta)$. Those estimates are easily obtained  by a  reduced run of our Gibbs algorithm outlined in Section 4.1 in which  $\theta$ is fixed. The degrees-of-freedom are set equal to $\nu^*=4$.

For any new value of $\theta$ at which the likelihood $f_\theta(\beta_{1:T})$ needs to be estimated, a new sequence of the SMC-IS densities (\ref{eq-app:IS-density}) are constructed by re-running the reduced Gibbs, which are then used to execute the SMC algorithm as described above.  The SMC log-likelihood estimates used to compute
the DIC criterion  according to Equation (17)
are obtained by setting the SMC sample size to  $L=200,000$ and using 100 cycles for the reduced Gibbs runs. The resulting log-likelihood estimates turned out to be very accurate as indicated by the MC-standard deviation obtained from rerunning the SMC algorithm under different seed. This  MC-standard deviation is about 0.0002 percent of the absolute value of the log-likelihood estimate.

\subsection*{A3: Implementation details for computing out-of-sample density and point forecasts}
{\bf MC-evaluation of the predictive density in Equation (19).} The predictive density for $y_{t+1}$ (as reproduced from Equation 19) is given by
\begin{align}\label{eq-app:predict_dens}
p(y_{t+1}|y_{1:t})&=
\int f_\theta(y_{t+1}|\beta_{t+1})f_\theta(\beta_{t+1}|\beta_t,H_{t+1})f_\theta(H_{t+1}|H_t)\\
&\qquad\qquad\qquad\qquad\qquad \times \pi(\beta_{1:t},H_{1:t},\theta|y_{1:t})d\beta_{1:t+1}dH_{1:t+1}d\theta. \nonumber
\end{align}
A brute-force MC estimate for  this  density  is  given by the sample mean of  $\{f_{\theta^{(j)}}(y_{t+1}|\beta_{t+1}^{(j)})\}_{j=1}^M$, where $\{ \beta_{t+1}^{(j)},\theta^{(j)} \}_{j=1}^M$ are simulated draws from  $f_\theta(\beta_{t+1}|\beta_t,H_{t+1})$$f_\theta(H_{t+1}|H_t)$$\pi(\beta_{1:t},H_{1:t},\theta|y_{1:t})$
based on Gibbs simulation from
 the period-$t$ posterior $\pi(\beta_{1:t},H_{1:t},\theta|y_{1:t})$. However, since under the fitted factor SSM model the measurement  density $f_\theta(y_{t+1}|\beta_{t+1})$ in Equation (\ref{eq-app:predict_dens}) is highly informative about the states $\beta_{t+1}$ such a brute approach  MC-approximation suffers
from a fairly low precision. In order to substantially increase the precision we exploit that under the factor SSM with its conditional linear Gaussian structure for $y_{t+1}$ and $\beta_{t+1}$ given $H_{t+1}$ (as assumed by
Equations 3 and 5), the states $\beta_{t+1}$ can be analytically integrated out. This simplifies  the predictive density integral in Equation (\ref{eq-app:predict_dens}) to
\begin{align}\label{eq-app:predict_dens_simp}
p(y_{t+1}|y_{1:t})&=
\int f_\theta(y_{t+1}|\beta_{t},H_{t+1})f_\theta(H_{t+1}|H_t) \pi(\beta_{1:t},H_{1:t},\theta|y_{1:t})d\beta_{1:t}dH_{1:t+1}d\theta,
\end{align}
where $f_\theta(y_{t+1}|\beta_{t},H_{t+1})$ is the density of a ${\cal N}(Z_{t+1}[\alpha+\Phi\beta_t] , Z_{t+1}H_{t+1}^{-1}Z_{t+1}'+\Sigma_y)$-distribution. Using this simplified representation, an  MC estimate of the predictive density obtains as
\begin{align}
\hat p(y_{t+1}|y_{1:t})= \frac{1}{M}\sum_{j=1}^M f_{\theta^{(j)}}(y_{t+1}|\beta_{t}^{(j)},H_{t+1}^{(j)}),
\end{align}
where $\{\beta_t^{(j)},H_{t+1}^{(j)},\theta^{(j)}\}_{j=1}^M$ are simulated draws from $f_\theta(H_{t+1}|H_t) \pi(\beta_{1:t},H_{1:t},\theta|y_{1:t})$.

{\bf MC-evaluation of the point and variance forecast in Equation (20).}
As MC-approximations for $\mbox{E}(\beta_{t+1}|y_{1:t},\theta)$ and $\mbox{Var}(\beta_{t+1}|y_{1:t},\theta)$  required to evaluate the forecasts in Equation (20), we use the sample mean and covariance matrix of  simulated values $\{\beta_{t+1}^{(j)}\}_{j=1}^M$  from the predictive density of the factors $f_\theta(\beta_{t+1}|y_{1:t})$. Draws from $f_\theta(\beta_{t+1}|y_{1:t})$ can be obtained by simulating
\begin{align}\label{eq-app:beta-pred-draws}
\beta_{t+1}^{(j)}\sim f_{\theta}(\beta_{t+1}|\beta_t,H_{t+1})f_{\theta}(H_{t+1}|H_t)\pi(\beta_{1:t},H_{1:t}|y_{1:t},\theta),\quad j=1,\ldots,M,
\end{align}
using reduced Gibbs runs to generate draws of the states from their conditional period-$t$  posterior  $\pi(\beta_{1:t},H_{1:t}|y_{1:t},\theta)$ in which $\theta$ is fixed at its posterior mean.

{\bf MC-evaluation of forecasts for the VaR of portfolios.} Under the factor SSM model in Equations (3)-(7) the period-($t+1$) log-return on a portfolio with a vector of weights $\omega$ is
\begin{align}
r_{t+1}^p = \omega'(y_{t+1}-y_t)=\omega'(Z_{t+1}\beta_{t+1}+\epsilon_{t+1} -y_t),
\end{align}
and its conditional density $f_\theta(r_{t+1}^p|\beta_{t+1},y_{1:t})$ given  $\beta_{t+1}$ and $y_{1:t}$  is the density of a ${\cal N}(\omega'[Z_{t+1}\beta_{t+1}-y_t], \omega'\Sigma_y\omega )$ distribution. The predictive density  for the portfolio returns written in terms of this  conditional density is
\begin{align}
f_\theta (r_{t+1}^p|y_{1:t}) = \int f_\theta(r_{t+1}^p|\beta_{t+1},y_{1:t}) f_\theta(\beta_{t+1}|y_{1:t})\beta_{t+1}.
\end{align}
Hence, a straightforward MC approximation to $f_\theta (r_{t+1}^p|y_{1:t})$  obtains by simulating
\begin{align}\label{eq-app:beta-pred-draws}
r_{t+1}^{p,(j)}\sim  f_\theta(r_{t+1}^p|\beta_{t+1},y_{1:t})f_{\theta}(\beta_{t+1}|y_{1:t}),\quad j=1,\ldots,M,
\end{align}
where draws from $f_{\theta}(\beta_{t+1}|y_{1:t})$ are generated according to Equation (\ref{eq-app:beta-pred-draws}). The $\alpha^*$-quantile of the empirical distribution of $\{r_{t+1}^{p,(j)}\}_{j=1}^M$  is  used as an MC estimate for the forecast  of the corresponding VaR.

\subsection*{A4: Prior information}
For the Bayesian analysis of our factor SSM model given in Equations (3)-(7) with the parameter restrictions described in Section 3.3 we assume the following prior distributions:
A flat prior is used for the log of the decay parameters, $(\ln \lambda_1,\ln \lambda_2)$ as well as for the degree-of-freedom parameter $\nu$. For the precision of the measurement errors we take a  Gamma prior with $1/\sigma_y^2\sim {\cal G}(\alpha^*,\beta^*)$, where  ${\cal G}(\alpha^*,\beta^*)$ denotes a Gamma distribution with parameters set equal to $\alpha^*=\beta^*=1$. A normal prior is used for  the VAR intercepts of the factors with $\alpha\sim{\cal N}(0,100^2 I_{m})$. As for the initial conditions of the state process, we assign to $\beta_0$ a Normal prior with $\beta_0\sim{\cal N}(0,1000 I_{m})$ and assume for $\Sigma_0$ a degenerate prior with $\Sigma_0=0.1^2I_{m}$.

For the restricted version of the factor SSM model  without stochastic volatility, i.e. $H_t^{-1}=\Sigma_0$,
we select for $\Sigma_0^{-1}$ a  Wishart prior with $\Sigma_0^{-1}\sim{\cal W}_m(m+10,S^*)$. The  scaling  matrix $S^*$ is selected such that the prior expectation  E$(\Sigma_0^{-1})=(m+1)S^*$ is equal to $0.15^2I_{m}$.

\subsection*{A5: Additional results}
Figure (A-1) plots estimates  of  the average  level  $\mbox{E}(y_{it})$ and variation $\mbox{Var}(y_{it})$ of the prices as given in Equation (8) for the contracts $i=1,4,16,24$. The estimates are those  obtained under the fitted  4-factor model with stochastic volatility (4F-SV) at the maturities $\tau_{it}$ observed for the first 43 trading days.

\section*{References}

\begin{small}
\begin{description}
\item[]Capp\'{e}, O., Godsill, S.J., and Moulines, E., 2007. An overview of existing methods and recent advances in sequential Monte Carlo. Proceedings of the IEEE 95, 899-924.\\[-1cm]
\item[]Doucet, A., and Johansen, A.M. (2009). A tutorial on particle filtering and smoothing: Fifteen years later.
       In: Crisan, D., Rozovskii, B. (eds), The Oxford Handbook of Nonlinear Filtering. Oxford University Press, 656-704.\\[-1cm]
\item[]Gordon, N.J., Salmond, D.J., and Smith, A.F.M., 1993. A novel approach to non-linear and non-Gaussian Bayesian state estimation. IEEE Proceedings F 140, 107-113.\\[-1cm]
\item[]Gupta, A.K., and Nagar, D.K., 2000. Matrix Variate Distributions. Chapman \& Hall/CRC, Boca Raton.\\[-1cm]
\item[]Robert, C.P., and Casella, G., 2004. Monte Carlo Statistical Methods. Springer, New York.\\[-1cm]
\item[]Windle, J., and Carvalho, C.M., 2014. A tractable state-space model for symmetric positive-definite matrices. Bayesian Analysis 9, 759-792.\\[-1cm]
\end{description}
\end{small}

\pagebreak

%
%
\begin{center}
\begin{figure}[!htb]
\centering
\includegraphics[width=0.7\textwidth,angle=0]{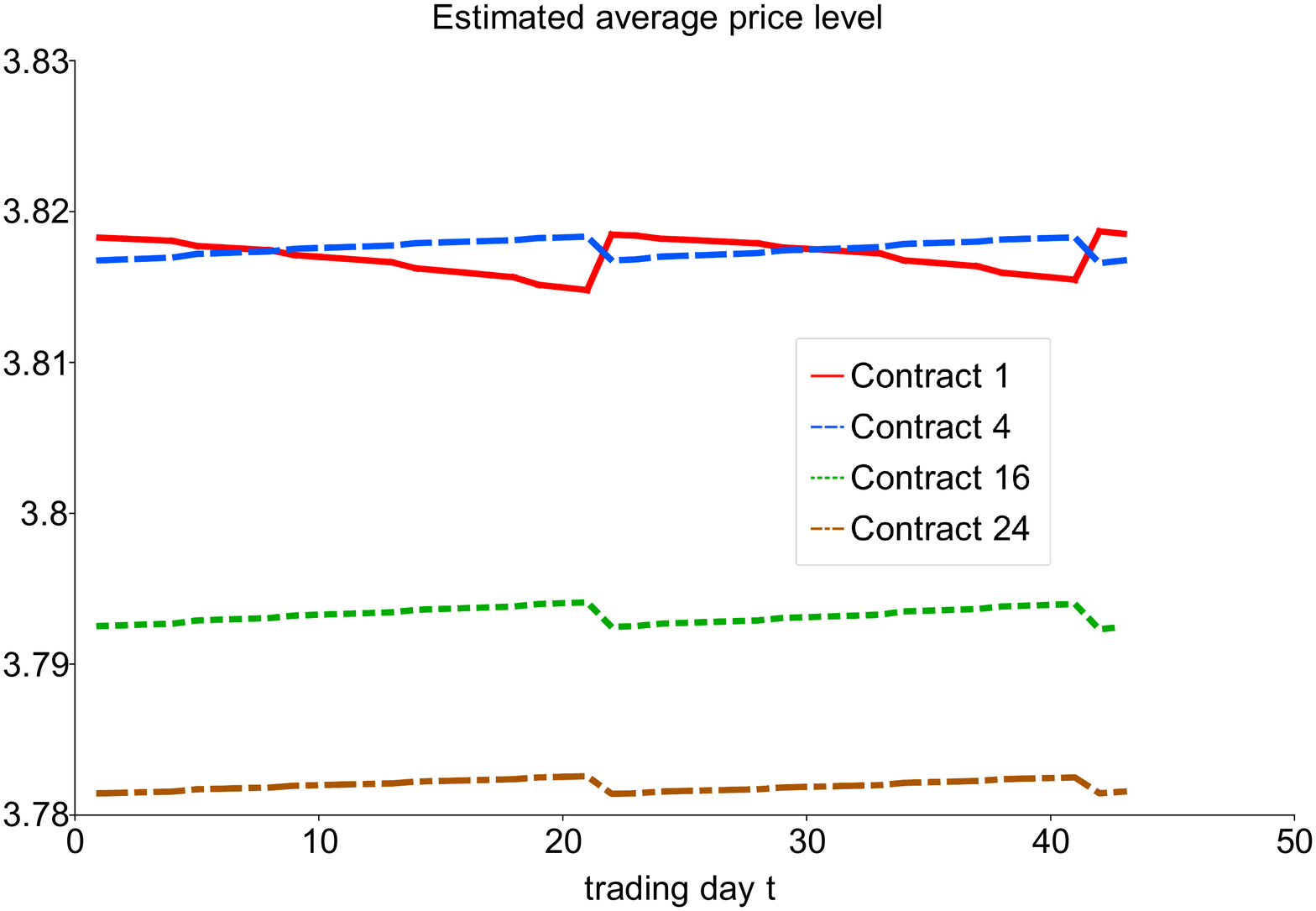}\\[-0.2cm]
\includegraphics[width=0.7\textwidth,angle=0]{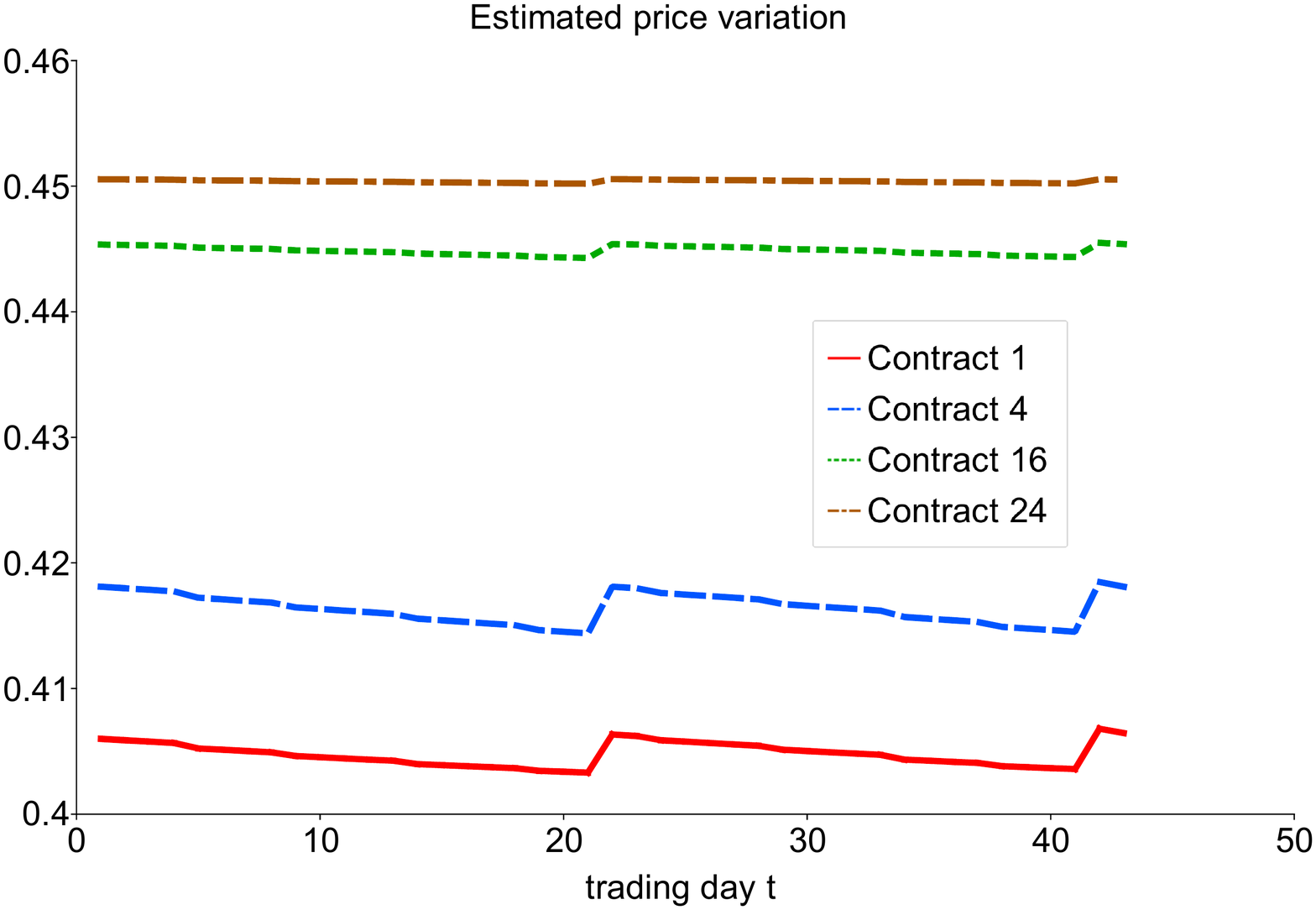}
\captionsetup{font=small}
\caption{Estimated average level  $\hat{\mbox E}(y_{it}) = \hat z_{it}' \hat {\mbox E} (\beta_t)$
(upper panel) and variation $\hat{\mbox V}{\mbox a}{\mbox r}(y_{it})= \hat z_{it}'\hat{{\mbox V}}{\mbox a}{\mbox r}(\beta_t) \hat z_{it}+ \hat\sigma_{y,i}^2$
(lower panel) of the  prices. $\hat z_{it}$  denotes the factor loadings in Equation (4) evaluated at the posterior estimates for $\lambda$.  The estimates $\hat {\mbox E} (\beta_t)$ and $\hat{{\mbox V}}{\mbox a}{\mbox r}(\beta_t)$ are computed as the sample mean and variance of the posterior means for $\{\beta_t\}_{t=1}^T$.    \label{fig:est_mean_var_data}}
\end{figure}
\end{center}


\end{document}